\documentclass[twocolumn, prapplied,secnumarabic, amssymb, nobibnotes, aps, superscriptaddress]{revtex4-1}
\usepackage{graphicx,subfigure,amsmath,mathrsfs}%
\usepackage{color,bm}

\usepackage[bookmarks=false]{hyperref}
\hypersetup{colorlinks=true,citecolor=blue,linkcolor=blue,urlcolor=blue,pdfstartview=FitH,bookmarksopen=true}
\begin{document}

	\title{Optimized Geometric Quantum Computation with mesoscopic ensemble of Rydberg Atoms}%

	\author{Chen-Yue Guo}%
	\affiliation{School of Physics, Zhengzhou University, Zhengzhou 450001, China}
	\author{L.-L. Yan}%
	\affiliation{School of Physics, Zhengzhou University, Zhengzhou 450001, China}
	\author{Shou Zhang}%
	\affiliation{Department of Physics, Yanbian University, Yanji 133002, China}
	\author{Shi-Lei Su}%
	\email{slsu@zzu.edu.cn}
	\affiliation{School of Physics, Zhengzhou University, Zhengzhou 450001, China}
	\author{Weibin Li}
	\affiliation{School of Physics and Astronomy, University of Nottingham, Nottingham NG7 2RD, United Kingdom}

	\date{\today}%
	\begin{abstract}
		We propose a nonadiabatic non-Abelian geometric quantum operation scheme to realize universal quantum computation with mesoscopic Rydberg atoms. A single control atom entangles a mesoscopic ensemble of target atoms through long-range interactions between Rydberg states. We demonstrate theoretically that both the single qubit and two-qubit quantum gates can achieve high fidelities around or above $99.9\%$ in ideal situations. Besides, to address the experimental issue of Rabi frequency fluctuation~(Rabi error) in Rydberg atom and ensemble, we apply the dynamical-invariant-based \emph{zero systematic-error sensitivity}~(ZSS) optimal control theory to the proposed scheme. Our numerical simulations show that the average fidelity could be $99.98\%$ for single ensemble qubit gate and $99.94\%$ for two-qubit gate even when the Rabi frequency of the gate laser acquires $10\%$ fluctuations. We also find that the optimized scheme can also reduce errors caused by higher-order perturbation terms in deriving the Hamiltonian of the ensemble atoms. To address the experimental issue of decoherence error between the ground state and Rydberg levels in Rydberg ensemble, we introduce a dispersive coupling regime between Rydberg and ground levels, based on which the Rydberg state is adiabatically discarded. The numerical simulation demonstrate that the quantum gate is enhanced. By combining strong Rydberg atom interactions, nonadiabatic geometric quantum computation, dynamical invariant and optimal control theory together, our scheme shows a new route to construct fast and robust quantum gates with mesoscopic atomic ensembles. Our study contributes to the ongoing effort in developing quantum information processing with Rydberg atoms trapped in optical lattices or tweezer arrays.

	\end{abstract}
	\maketitle

	\section{introduction}

	In recent years, Rydberg atoms have been extensively used in the study of quantum information processing due to their unique properties~\cite{djp2000,mmr2001,mtk2010,Saffman_2016}.  Rydberg atoms can be trapped in optical lattices or tweezer arrays~\cite{anderson_trapping_2011,mmtm2015,tmtt2015,labuhn_tunable_2016,bernien_probing_2017}, and moreover have long lifetimes~\cite{tfg1984,mmr2001}. Once excited, strong dipole-dipole or van der Waals interactions beween Rydberg atoms induce a energy shift, which hinders more than one atom from being excited to Rydberg states in an ensemble of atoms~\cite{dsjs2004,ruvb2007,wjam2009,urban_observation_2009,comparat2010dipole,dudin_observation_2012,ya2012,Saffman_2016}, leading to the interaction induced blockade effect. Directly using the strong interaction, early proposals~\cite{djp2000,mmr2001} have shown that two-qubit quantum gates can be realized with Rydberg atoms. These seminar works have triggered a growing interest in the study of Rydberg gates with a number of proposals of different gate schemes~\cite{Protsenko2002,Xia2013,Muller2014,Goerz2014,Theis2016,Shi2018,Huang2018,Sun2020,dressing2015,dressing2020,Petrosyan2014,Su2016,*Su2017,*Su2017a,*Su2018,*Su2020,Rao2014,Petrosyan2017,Beterov2018,*Beterov2018a,Shi2017,*Shi2017a,*Shi2018a,*Shi2019,beterov2020application,ljd2015,ekm2007,Saffman2008,mihhp2009,ybkc2010,Wu2010,KlausRydberggeometric,zb2012,imed2013,zhaoensemble,kang2018,Zhaopeizi2017,wuhuaizhi2017,Shen:19,Liao2019,Qi:20}. According to the cause of conditional operations, we have dynamical gates~\cite{Protsenko2002,ekm2007,Saffman2008,mihhp2009,ybkc2010,Wu2010,Xia2013,Muller2014,Goerz2014,Theis2016,Petrosyan2014,Petrosyan2017,Rao2014,Su2016,*Su2017,*Su2017a,*Su2018,*Su2020,Beterov2018,*Beterov2018a,Shi2017,*Shi2017a,*Shi2018a,*Shi2019,Shi2018,Huang2018,Sun2020,beterov2020application,ljd2015,dressing2015,dressing2020} and geometric gates~\cite{KlausRydberggeometric,zb2012,imed2013,Zhaopeizi2017,wuhuaizhi2017,kang2018,zhaoensemble,Shen:19,Liao2019,Qi:20}. Depending on numbers of Rydberg states populated in gate operations, these schemes can be divided into blockade~\cite{djp2000,Protsenko2002,ekm2007,Saffman2008,mihhp2009,ybkc2010,Wu2010,Xia2013,Muller2014,Goerz2014,Theis2016,Shi2018,Huang2018,Sun2020,KlausRydberggeometric,zb2012,imed2013,Zhaopeizi2017,wuhuaizhi2017,kang2018,zhaoensemble,Shen:19,Liao2019,Qi:20}, Rydberg dressing~\cite{djp2000,Muller2014,dressing2015,dressing2020}, antiblockade~\cite{djp2000,Petrosyan2014,Su2016,*Su2017,*Su2017a,*Su2018,*Su2020}, towards two-excitation Rydberg state~\cite{Rao2014}, dipole-dipole resonant interaction~\cite{Petrosyan2017}, and F\"{o}rster resonance~\cite{Beterov2018,*Beterov2018a,beterov2020application}. From the number of atoms involved in quantum computation, these schemes can be classified as single atoms scheme~\cite{Protsenko2002,Xia2013,Muller2014,Goerz2014,Theis2016,dressing2015,dressing2020,Petrosyan2014,Petrosyan2017,Rao2014,Su2016,*Su2017,*Su2017a,*Su2018,*Su2020,Beterov2018,*Beterov2018a,Shi2017,*Shi2017a,*Shi2018a,*Shi2019,Shi2018,Huang2018,Sun2020,Zhaopeizi2017,wuhuaizhi2017,kang2018,Shen:19,Liao2019,Qi:20} or mesoscopic ensemble scheme~\cite{mmr2001,ekm2007,Saffman2008,KlausRydberggeometric,mihhp2009,ybkc2010,Wu2010,zb2012,imed2013,ljd2015,zhaoensemble,beterov2020application}. Recent experiments have demonstrated quantum logic gates between single Rydberg atoms~\cite{lex2010,xla2010,tac2010,kmt2015,ypx2017,Picken_2018,saa2018,zeng_entangling_2017,levine2019parallel,graham2019rydberg,Omran570}.

Besides schemes based on strong two-body interactions, one can construct robust quantum logic gates~\cite{ZANARDI199994, Jones2000,Duanluming,Wangxiangbin,Zhushiliang} by applying geometric quantum operations~\cite{Berryphase,Wilczek}. In this family of quantum computation, nonadiabatic holonomic quantum computation~\cite{edlb2012,gjde2012}~(NHQC)~based on nonadiabatic non-Abelian geometric phases~\cite{j1988} has the advantage to generate robust geometric phases against certain parameter fluctuations, and does not need to satisfy the adiabatic condition. Then, the NHQC was studied in the noiseless subsystems to suppress the detrimental effects induced by the coupling between system and environment~\cite{Zhang2014}. After this, the researchers have also studied how to implement an arbitrary nonadiabatic holonomic gate via a single-shot method~\cite{Xu2015,SJOQVIST201665,Zhao2017}. These studies further enrich the NHQC dynamics and make it being a hot research topic in quantum computation. Experimentally, NHQC has been demonstrated with nuclear magnetic resonance~\cite{ggg2013,hyg2017,Zhu2019}, superconducting circuits~\cite{ajkm2013,ywyx2018,saj2018,dmga2019,ylx2019,zptl2019}, and nitrogen-vacancy centers in diamond~\cite{cwlw2014,sasg2014,ynrh2017, bpvf2017,kkyh2018,ntts2018}.
Besides, the schemes~\cite{Zhang2015,*Liang2016,*Song2016,*WuandSu2019} combining geometric quantum computation with shortcut-to-adiabaticity~\cite{Muga2020}, called NHQC+, have also been studied theoretically~\cite{lsx2019} and experimentally~\cite{ylx2019}. In Ref.~\cite{KlausRydberggeometric}, it was proposed to realize slow geometric phase gates using Rydberg atoms through adiabatic passage.

In Rydberg atom experiments, gate operations based on existing schemes suffer from many errors. When coupling ground and Rydberg states, laser power and frequency can fluctuate, affecting, for example, Rabi frequencies. When excited, Rydberg atoms are typically influenced by finite lifetimes, leading to decoherence.  As a result, the parameter fluctuation and decoherence can decreases gate fidelities. In other words, currently, the low fidelity in Rydberg atom system are mainly caused by fluctuations of Rabi frequency and decoherence.  To address these issues, in viewing the advantages seen in other systems, we use nonadiabatic holonomic quantum computation and optimal control theory between single control atom and a target mesoscopic Rydberg atom ensemble~(MRAE) to suppress the imperfections induced by Rabi frequency fluctuations. Besides, we also consider a dispersion mechanism to reduce the effect of decoherence. 

When there is only one Rydberg excitation, a MRAE can be described by a superatom, consisting of a collective groundstate (all atoms are in the groundstate) and collective excited state where the Rydberg excitation is shared by all the atoms. Laser manipulation of the superatom is fast due to the collective coupling. Therefore this scheme combines the controllability of MRAEs and the feature of geometric-phase-based NHQC~\cite{edlb2012,gjde2012}. To further improve the gate fidelity and robustness, we use the invariant-based inverse engineering method to redesign the gate pulses~\cite{lsx2019,ylx2019}. This approach is compatible with the \emph{zero systematic-error sensitivity}~(ZZS)~\cite{Ruschhaupt_2012} optimal control method. This brings additional advantages such that the gate is robust even in the presence of systematic errors. And the error induced by ignoring higher-order perturbation terms could also be decreased. The improvement is dramatic in the operation of quantum gates. Through numerical calculations, we demonstrate that the fidelity of single-control-qubit and single-ensemble-qubit~(i.e., MRAE) of the NHQC scheme are about 0.9999 and 0.999, respectively, and the fidelity of two-qubit controlled-NOT gate is 0.9971 [ensemble atom number $N=4(8)$], in ideal situations. For the optimized NHQC+ gates under static systematic error, the scheme can still maintain the higher fidelity (for single-ensemble-qubit, 0.9998; for two-qubit, higher than 0.999) even when the Rabi frequency fluctuates as large as 10\% around the mean value.

The schemes studied in this work have the following fascinating features. Firstly, the target qubit is MRAE, which can be coupled strongly by lasers, hence speeding up the gate preparation. Secondly, the NHQC operations constructed in MRAE, which combine the advantages of geometric phase and Rydberg atom together, show very high fidelities. Thirdly, the robustness and fidelity are further optimized based on dynamical invariant and ZZS optimal control theory. These features demonstrate that the present mesoscopic Rydberg quantum computation scheme is robust. It has the potential to overcome technological challenges due to laser fluctuations and decoherence of cold Rydberg atom systems. Our study will benefit to current experimental efforts in building robust and fast quantum gates with Rydberg atoms.
	
The structure of the paper is as follows: In Sec.~\ref{s2n}, we describe the Hamiltonian for different elements in the gate, and master equation that governs dynamics of the system. In Sec.~\ref{s2}, we introduce requirements of the NHQC gate scheme and demonstrate how to realize high fidelity one-qubit and two-qubit NHQC gates with Rydberg atom ensembles. In Sec.~\ref{s20}, we show how to implement NHQC+ gates via inverse-engineering-based optimal control. We show that the gate schemes is robust against parameter fluctuations. In Secs.~\ref{s5} and \ref{s6}, discussion and conclusion are given, respectively.

	\begin{figure}
		\includegraphics[width=\linewidth]{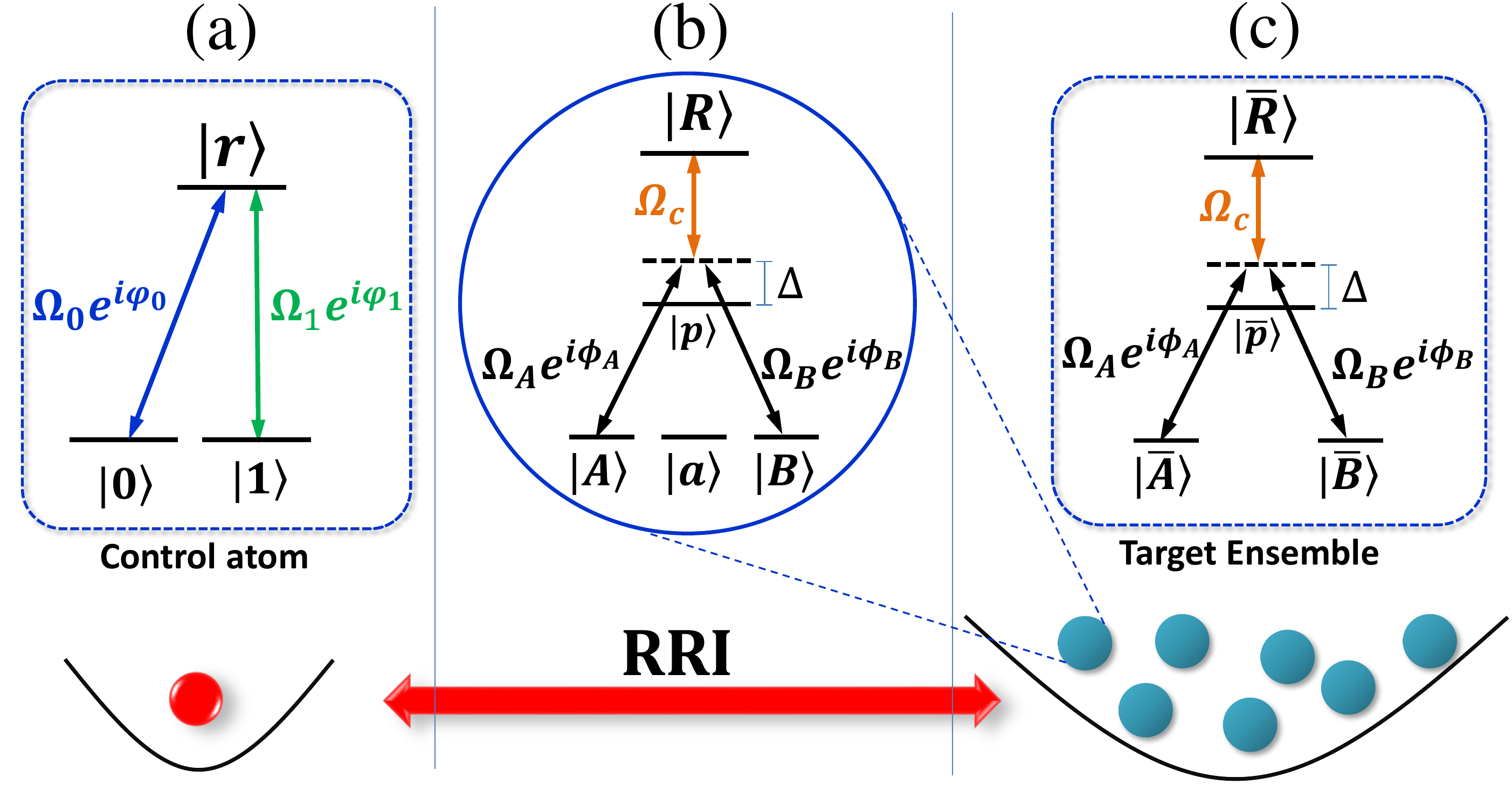}\\
		\caption{Configuration of the scheme. (a) Single control atom. $|0\rangle$ and $|1\rangle$ are two ground states encoding quantum information while $|r\rangle$ denotes Rydberg state. The ground state $|0\rangle$ and $|1\rangle$ is resonantly coupled to the Rydberg state $|r\rangle$ by complex Rabi frequency $\Omega_0e^{i\varphi_0}$ and $\Omega_{1}e^{i\varphi_1}$~(where $\varphi_ j$ is the laser phase with $j=0,~1$), respectively. (b) Single ensemble atom. $|A\rangle$, $|a\rangle$, and $|B\rangle$ are three ground states, where $|a\rangle$ denotes auxiliary state. $|p\rangle$ and $|R\rangle$ denote the intermediate state and Rydberg state, respectively. $|A\rangle$~($|B\rangle$) are off-resonantly coupled to $|p\rangle$ with detuning $\Delta$ and complex Rabi frequency $\Omega_{A}e^{i\phi_A}(\Omega_{B}e^{i\phi_B})$, in which $\phi_A$ and $\phi_{B}$ being the laser phase. Then if the control atom is not excited, the two-photon process form $|A\rangle~(|B\rangle)$ to $|R\rangle$ is feasible. Inversely, the two-photon process would be inhibited due to the RRI. (c) Equivalent energy level diagram of ensemble qubit. This configuration can be constructed based on the experimental parameters in Refs.~\cite{mmtm2015} and ~\cite{lex2010} when the atom number of ensemble is less than 10~\cite{mmtm2015}.}\label{f001}
	\end{figure}
\section{Model and holonomic dynamics}\label{s2n}

In this section, we introduce basic elements, including single control atom, single atoms in the mesoscopic ensemble, and single-ensemble-qubit, in realizing single and two-qubit gates. We will present Hamiltonians that govern their dynamics and respective master equations in the presence of dissipation.
\subsection{Single control atom}
The level scheme of the control atom is shown in Fig.~\ref{f001}(a). Quantum information is encoded in the two ground states $|0\rangle$  and $|1\rangle$. They are coupled to a Rydberg state $|r\rangle$ resonantly with complex Rabi frequencies $\Omega_0e^{i\varphi_0}$ and $\Omega_1e^{i\varphi_1}$ with $\Omega_j$ and $\varphi_j~(j=0,~1)$ to be the amplitude and phase of the coupling.  Under the rotating wave approximation, the Hamiltonian of the control atom is written as
	\begin{eqnarray}\label{e01n}
	\hat{H}_{c}=\frac{\Omega}{2}\left[\sin\frac{\theta}{2}e^{i\varphi_{0}}|0\rangle\langle r|+\cos\frac{\theta}{2}e^{i\varphi_{1}}|1\rangle\langle r|\right]+ {\rm H.c.},
	\end{eqnarray}
where $\Omega=\sqrt{\Omega_{0}^{2}+\Omega_{1}^{2}}$. And $\theta$ is the rotation angle satisfing $\tan(\theta/2)=\Omega_{0}/\Omega_{1}$ and is kept as a constant.
By using the dressed states $|d\rangle=\cos(\theta/2)|0\rangle-\sin(\theta/2)e^{-i(\varphi_0-\varphi_1)}|1\rangle$ and $|b\rangle=\sin(\theta/2)e^{i(\varphi_0-\varphi_1)}|0\rangle+\cos(\theta/2)|1\rangle$, Eq.~(\ref{e01n}) can be rewritten in a compact form, $\hat{H}_{c} = \Omega/2e^{i\varphi_1}|b\rangle\langle r| + {\rm H.c.}$

\subsection{Single atom in the ensemble}\label{s2.2.2}
The level scheme of a single atom in a MRAE is shown in Fig.~\ref{f001}(b). Each atom will have three ground states $|A\rangle$, $|a\rangle$ and $|B\rangle$. State $|A\rangle$~($|B\rangle$) is off-resonantly coupled to the intermediate state $|p\rangle$ with detuning $-\Delta$ and complex Rabi frequency $\Omega_{A}e^{i\phi_A}(\Omega_{B}e^{i\phi_B})$
with $\phi_A~(\phi_B)$ being the laser phase. State $|p\rangle$ off-resonantly couples to a Rydberg state $|R\rangle$ with Rabi frequency $\Omega_C$ and detuning $\Delta$. The  Hamiltonian of the atom reads
\begin{equation}\label{e02n}
\hat{H}_{e} = \frac{1}{2}e^{i\Delta t}\left[\Omega'e^{i\phi_{B}}|\mathcal{B}\rangle\langle p|+{\Omega_{C}}|R\rangle\langle p|\right]+ {\rm H.c.},
\end{equation}
in which $\Omega'=\sqrt{\Omega_{A}^{2}+\Omega_{B}^{2}}$, $\tan(\vartheta/2) =\Omega_{A}/\Omega_{B}$, $|\mathcal{B}\rangle=\sin(\vartheta/2)e^{i(\phi_{A}-\phi_{B})}|A\rangle+\cos(\vartheta/2)|B\rangle$, $|\mathcal{D}\rangle = \cos(\vartheta/2)|A\rangle-\sin(\vartheta/2)e^{-i(\phi_{A}-\phi_{B})}|B\rangle$ denotes dark state that decoupled from the dynamics. Besides, $\vartheta$ should be kept as constant during the gate operation.
Under the condition of large detuning $\Delta\gg\{\Omega',~\Omega_{C}\}$, state $|p\rangle$ can be adiabatically eliminated from the dynamics by applying the second-order perturbation theory~\cite{dj2007} while high-order terms are neglected. This leads to an effective Hamiltonian  as
\begin{equation}\label{e03n}
\hat{H}_{\rm s} =\hat{H}_{0}+\hat{H}_{I}
\end{equation}
with $\hat{H}_{0}=\frac{\Omega'^{2}}{4\Delta}|\mathcal{B}\rangle\langle\mathcal{B}|
+\frac{\Omega_{C}^{2}}{4\Delta}|R\rangle\langle R|$ and $\hat{H}_{I}=\frac{\Omega'\Omega_{C}e^{i\phi_{B}}}{4\Delta}|\mathcal{B}\rangle\langle R|+ {\rm H.c.}$.
The stark shifts given in $\hat{H}_0$ can be canceled out by tuning the lasers~\cite{note}.
With these considerations, Eq.~(\ref{e03n}) is simplified to be
\begin{equation}\label{e04n}
\hat{H}_{\rm s} =\frac{\Omega'\Omega_{C}e^{i\phi_{B}}}{4\Delta}|\mathcal{B}\rangle\langle R|+ {\rm H.c.}
\end{equation}

\subsection{Single-ensemble-qubit}\label{s2.1.3}
Before introducing the basic model of single-ensemble-qubit of MRAE, we should point out that, although the energy level configuration of our scheme is inspired by Ref.~\cite{mihhp2009}, the laser parameter range of $\Omega_{C}$ and ensemble qubit encoding method are very different. In Ref.~\cite{mihhp2009}, Electromagnetically Induced Transparency~(EIT)~regime is considered thus $\Omega_{C}\gg\{\Omega_{A},~\Omega_{B}\}$ should be fulfilled, which is not the case of our scheme~(see Sec.~\ref{s2.2.2}). The ensemble qubit of Ref.~\cite{mihhp2009} is encoded as $|A^N\rangle\equiv\otimes_{k=1}^{N}|A\rangle_{k}$ and $|B^N\rangle\equiv\otimes_{k=1}^{N}|B\rangle_{k}$ (footnote \emph{k} denotes the \emph{k}-th atom), respectively. Then, the NOT gate of the ensemble qubit $\hat{\sigma}_{x,L}$ is equivalent to the direct multiplication of the NOT gate of each ensemble atom,  i.e., $\hat{\sigma}_{x,L}=\otimes_{k=1}^{N}\hat{\sigma}_{x,k}$, where $\hat{\sigma}_{x,k}$ denotes NOT operation on the \emph{k}-th atom. Nevertheless, for more general operations $\hat{\mathcal M}$, one can verify that perform $\hat{\mathcal M}$ on ensemble qubit is not equal to perform $\hat{\mathcal M}$ on each ensemble atom.

Here we will apply a different encoding protocol to construct the controlled-universal operations. Inspired by Refs.~\cite{imed2013, mmtm2015}, we consider the ensemble qubit of \emph{N}-atom MRAE as
\begin{equation}\label{e05n}
  |\overline{\zeta}\rangle = \frac{1}{\sqrt{N}} \sum_{l=1}^{N}|a\rangle_{1}|a\rangle_{2}\cdots|\zeta\rangle_{l}\cdots|a\rangle_{N},
  \end{equation}
  where $\zeta\in\{A,~B, ~p,~R,~\mathcal{B},~\mathcal{D}\}$. These states can be generated through the Rydberg blockade, as described in Appendix~\ref{s7}. For $N$ atoms, the total Hamiltonian of the MRAE is given by
  \begin{equation}\label{e06n}
  \hat{\mathcal{H}}_{\rm eff} = \sum_{l=1}^{N}\frac{\Omega'\Omega_{C}e^{i\phi_{B}}}{4\Delta}|\mathcal{B}\rangle_{l}\langle R|+ {\rm H.c.}
  \end{equation}
  Using the ensemble qubit state defined in Eq.~(\ref{e05n}), the total Hamiltonian can be rewritten as
  \begin{equation}\label{e07n}
  \hat{\mathcal{H}}_{\rm eff} =\frac{\Omega'\Omega_{C}e^{i\phi_B}}{4\Delta}|\overline{\mathcal{B}}\rangle\langle \overline{R}|+ {\rm H.c.},
  \end{equation}
where $|\overline{\mathcal{B}}\rangle=\sin(\vartheta/2)e^{i(\phi_{A}-\phi_{B})}|\overline{A}\rangle+\cos(\vartheta/2)|\overline{B}\rangle$. The collective dark state $|\overline{\mathcal{D}}\rangle=\cos(\vartheta/2)|\overline{A}\rangle-\sin(\vartheta/2)e^{-i(\phi_A-\phi_B)}|\overline{B}\rangle$ is decoupled from the system dynamics.

\subsection{Master equation and average fidelity}
Taking into account of the spontaneous decay in state $|p\rangle$ and Rydberg states, dynamics of the system is governed by the master equation
\begin{equation}\label{e16j}
\dot{\hat{\rho}}=-i[\hat{H},\hat{\rho}]+\sum_{j}\hat{L}_{j}[\hat{\rho}]+\sum_{i=1}^{N}\sum_{g,e}\hat{L}_{i,g,e}[\hat{\rho}].
\end{equation}
in which
$\hat{L}[\hat{\rho}]=\hat{L}\hat{\rho}\hat{L}^\dagger-\frac{1}{2}(\hat{L}^\dagger\hat{L}\hat{\rho}+\hat{\rho}\hat{L}^\dagger\hat{L})$, $\hat{L}_{j}=\sqrt{\gamma_{r}/2}|j\rangle\langle r| (j=0,~1)$ describes the spontaneous emission process of control atom with rate $\gamma_{r}$. Also, $L_{i,g,e}=\sqrt{\gamma_{e}/3}|a_{1}...g_{i}...a_{N}\rangle\langle a_{1}...e_{i}...a_{N}|$ denotes the spontaneous emission process from the state $|e\rangle (e = R,~p$) to the ground states $|g\rangle (g = A,~B,~a$) of the \emph{i}-th atom with rate $\gamma_{e}$. And $\hat{H}$ denotes the Hamiltonian of the system.
We numerically solve the master equation~(\ref{e16j}) with given sets of parameters by using forth-order Runge-Kutta method. With the solution at hand, we can calculate dynamical evolution of the initial state and fidelities of different gates.

The performance of quantum gates is measured by evaluating the average fidelity, which provides a better measure than considering special states. In this work, the average fidelity is given by~\cite{NIELSEN2002249,White:07}
\begin{eqnarray}\label{e17j}
F(\hat{U},\varepsilon)=
\frac{\sum_{j}{\rm tr}[\hat{U}\hat{U}_{j}^\dagger\hat{U}^\dagger\varepsilon(\hat{U}_{j})]+d^2}{d^2(d+1)},
\end{eqnarray}
in which $\hat{U}$ is the ideal quantum logic gate. For single qubits, $\hat{U}$ represents the truth table of the NOT, Hadamard, and $\pi$-phase gates. For two qubits, $\hat{U}$ represents the corresponding controlled NOT, Hadamard, and Z gates. $\varepsilon$ is the trace-preserving quantum operation, and $\hat{U}_{j}$ is the tensor of Pauli matrices $\hat{I},\hat{\sigma}_{x},\hat{\sigma}_{y},\hat{\sigma}_{z}$ for single-qubit or $\hat{I}\hat{I},\hat{I}\hat{\sigma}_{x},\hat{I}\hat{\sigma}_{y}...\hat{\sigma}_{z}\hat{\sigma}_{z}$ for two-qubit quantum gate. And, $d=2^n$ with \emph{n} denoting the number of qubit in the quantum logic gate.

\section{NHQC gates}\label{s2}
In this section, we will first introduce the holonomic constraints in quantum dynamics in implementing NHQC. Then we will demonstrate how to realize single and two-qubit NHQC gates with the Rydberg atom setting.
\subsection{Requirements of the NHQC scheme}
Lets consider a quantum system with Hamiltonian $H(t)$, in which the evolution operator reads $U(t,0)=\mathcal{T}\exp[-i\int_{0}^{t}H(t')dt']$ and a time-dependent \emph{L}-dimensional subspace $S(t)$ spanned by the orthogonal basis vector $\{|\phi_k(t)\rangle\}(k=1,\cdots,L)$ which satisfies $i|\dot{\phi}_k(t)\rangle=H(t)|\phi_k(t)\rangle$ at each instant \emph{t}. It has been said in the scheme~\cite{gjde2012}, the unitary transformation is holonomy matrix acting on the \emph{L}-dimensional subspace $S(0)$ spanned by $\{|\phi_k(0)\rangle\}(k=1\cdots L)$ if $|\phi_k(t)\rangle$ satisfies the conditions: i) $\sum_{k=1}^{L}|\phi_k(\tau)\rangle\langle \phi_k(\tau)|=\sum_{k=1}^{L}|\phi_k(0)\rangle\langle \phi_k(0)|$ and ii) $\langle \phi_k(t)|H(t)|\phi_l(t)\rangle=0$, with $\{k,~l\}=1,\cdots,L$.
Where the condition i) shows the subspace undergoes a cyclic evolution; and ii) is the parallel-transport condition. Similar conditions are also given in Ref.~\cite{edlb2012}.

\subsection{Single-qubit gate}\label{s2.1}
Since the control qubit is single Rydberg atom and the target qubit is a MRAE, we will consider the single-qubit gates for control and target qubit, respectively. We will show that their dynamics fulfill the holonomic constraints given above.

\subsubsection{Single control atom}
For a single control atom, Hamiltonian~(\ref{e01n}) reads
	\begin{eqnarray}\label{e01}
   \hat{H}_{c} = \Omega/2|b\rangle\langle r| + {\rm H.c.},
	\end{eqnarray}
	where $|b\rangle=\sin(\theta/2)e^{i\varphi}|0\rangle+\cos(\theta/2)|1\rangle$ and $|d\rangle=\cos(\theta/2)|0\rangle-\sin(\theta/2)e^{-i\varphi}|1\rangle$. In writing the Hamiltonian, we have set $\varphi_0 = \varphi$ and $\varphi_1 = 0$.
	The evolution operator of the atom is $\hat{U}=e^{{-i}{\int_{0}^{T}\hat{H}_{c}(t)\,dt}}$. If the initial state is in the ground-state subspace and the laser pulse fulfills $\int_{0}^{T}\Omega(t) \emph{d}t=2\pi$, the evolution operation becomes $\hat{U}=-|b\rangle\langle b|+|d\rangle\langle d|$ at the gate time $T$. In the bare basis $\{|0\rangle,~|1\rangle\}$, the evolution operator can be written as
	\begin{equation}\label{e02}
	\hat{U}(\theta,~\varphi)=\left(
	\begin{array}{cc}
	\cos\theta    & -\sin\theta e^{i\varphi} \\

	-\sin\theta e^{-i\varphi}   & -\cos\theta \\
	\end{array}
	\right).
	\end{equation}

	One can adjust parameters $\theta$ and $\varphi$ independently to achieve the expected single-qubit NHQC gate. For instance, $\{\theta, \varphi\}$ equals $\{-\pi/2, 0\}$ for NOT gate, $\{-\pi/4, 0\}$ for Hadamard gate, and $\{0, 0\}$ for $\pi$ phase gate, respectively.	With these choices, we can check the conditions of the NHQC are met. Firstly, the condition of cyclic evolution is satisfied: $|d\rangle\rightarrow|d\rangle,|b\rangle\rightarrow-|b\rangle$. Secondly, the condition of parallel-transport is also satisfied: $\langle\psi_k{(t)}| \hat{H}_{c}|\psi_l{(t)}\rangle=\langle \psi_k{(0)}| U^\dagger \hat{H}_{c}U|\psi_l{(0)}\rangle=0$ with $\{|\psi_k{(0)}\rangle,|\psi_l{(0)}\rangle\}\in\{|0\rangle,|1\rangle\}$.

	\subsubsection{Single ensemble qubit}\label{s2.1.2}
For convenience, we will set $\phi_{A} = \phi$ and $\phi_B = 0$ in the Hamiltonian of the single ensemble qubit. Then Hamiltonian~(\ref{e07n}) becomes
	\begin{equation}\label{e010a}
	\hat{\mathcal{H}}_{\rm eff} =\frac{\Omega'\Omega_{C}}{4\Delta}|\overline{\mathcal{B}}\rangle\langle \overline{R}|+ {\rm H.c.},
	\end{equation}
	where $|\overline{\mathcal{B}}\rangle=\sin(\vartheta/2)e^{i\phi}|\overline{A}\rangle+\cos(\vartheta/2)|\overline{B}\rangle$ and $|\overline{\mathcal{D}}\rangle=\cos(\vartheta/2)|\overline{A}\rangle-\sin(\vartheta/2)e^{-i\phi}|\overline{B}\rangle$. If ${\int_{0}^{T}\Omega'\Omega_{c}/(2\Delta) dt = 2\pi}$ is fulfilled, the evolution operator $\hat{\mathcal{U}}=e^{{-i}{\int_{0}^{T}\hat{\mathcal{H}}_{\rm eff}(t) dt}}$ becomes $\hat{\mathcal{U}}=-|\overline{\mathcal{B}}\rangle\langle\overline{\mathcal{B}}|+|\overline{\mathcal{D}}\rangle\langle\overline{\mathcal{D}}|$ at time $T$. It can be re-expressed in the matrix form as
	\begin{equation}\label{e011}
	\hat{\mathcal{U}}(\vartheta,~\phi)=\left(
	\begin{array}{cc}
	\cos\vartheta    & -\sin\vartheta e^{i\phi} \\
	-\sin\vartheta e^{-i\phi}   & -\cos\vartheta \\
	\end{array}
	\right)
	\end{equation}
	in the basis \{$|\overline{A}\rangle,~|\overline{B}\rangle$\}.

Desired single-ensemble-qubit gates can be realized by choosing suitable parameters. For example, different universal gates can be achieved by choosing corresponding values of $\vartheta$ and $\phi$. We can choose $\{\vartheta, \phi\}$ =$\{-\pi/2, 0\}$ for NOT gate, $\{-\pi/4, 0\}$ for Hadamard gate, and $\{0, 0\}$ for $\pi$ phase gate, respectively. Similar to the single control atom, one can check that the conditions of NHQC are satisfied for the single ensemble qubit.

\begin{figure}
		\includegraphics[width=\linewidth]{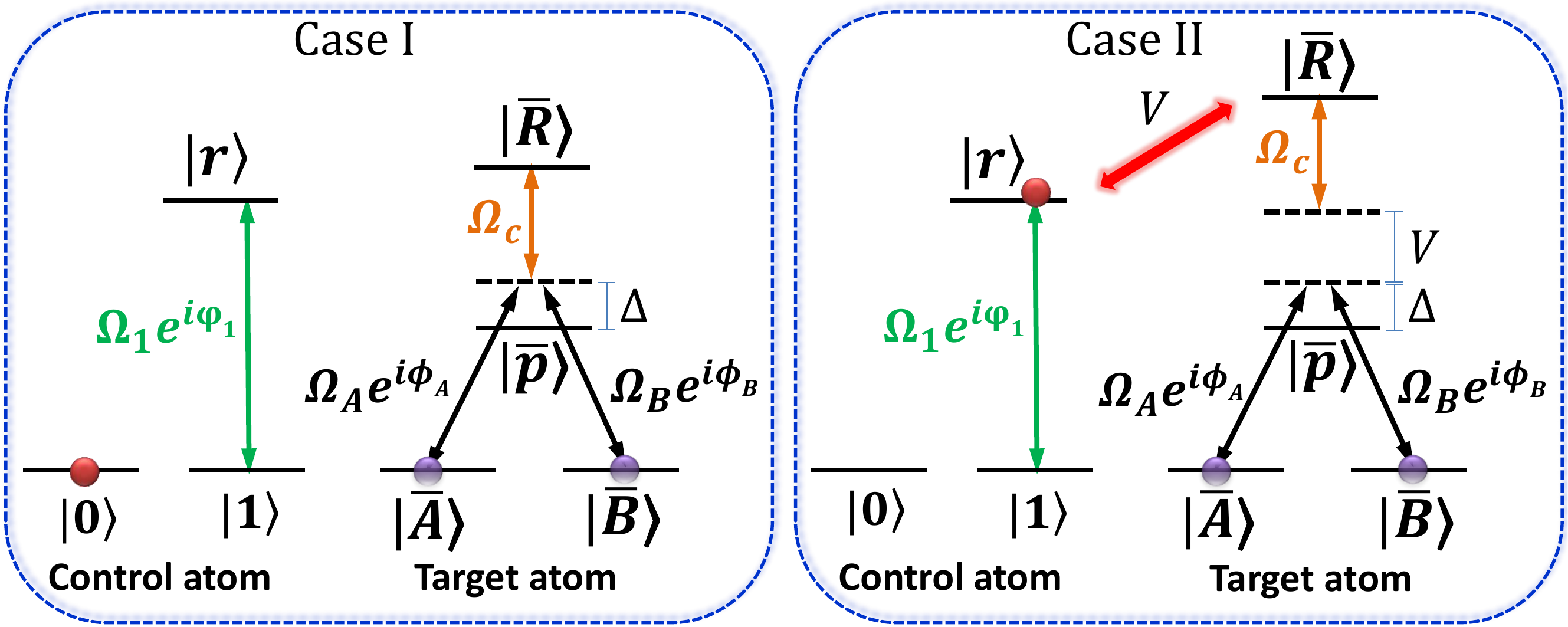}\\
		\caption{(Color online) Two cases of step~(ii). The two-photon process of the target ensemble-qubit would be inhibited or not conditioned on the state of the control atom. \emph{V} denotes the RRI strength.}\label{f002}
	\end{figure}
\subsection{Two-qubit gate}\label{s3}
The two-qubit gates are realized in three steps.

Step~(i):  Set $\Omega_{0}=0$ and $\varphi_{1}=0$, and then excite the control atom.
As shown in Fig.~\ref{f002}, the Hamiltonian of the control atom then reads
	\begin{equation}\label{e04}
	\hat{H}_{c} =\frac{\Omega_{1}}{2}|1\rangle\langle r|+ {\rm H.c.}
	\end{equation}
When $\int_{0}^{\tau_{c}/2}\Omega_{1}(t)dt=\pi$, the condition $|1\rangle\rightarrow -i|r\rangle$ can be achieved. One can check that in this step the condition ii) for NHQC is satisfied. And we would show that the condition i) of NHQC for control atom would be satisfied after considering all steps.

	Step (ii): Turn on the lasers on the ensemble atoms. We divide this step in two cases. In Case I, the control atom is not excited in step~(i). There is no RRI when the ensemble is illuminated.
	We then perform the same operations shown in Sec.~\ref{s2.1.2}, and we would get the result same as Eq.~(\ref{e011}) for the ensemble qubit and the NHQC conditions are also satisfied. In Case II, the control atom is excited to the Rydberg state $|r\rangle$ after step~(i). There will be a shift on the energy of Rydberg state of ensemble atoms via the RRI~[See Fig.~\ref{f002}]. The energy shift induced by the interstate interaction lifts the two-photon-resonance condition, which inhibits the operations on the ensemble. The Hamiltonian of single ensemble atom in the rotation frame can be written as
	\begin{eqnarray}\label{e10j}
	\hat{H}_{e} =&& \frac{1}{2}e^{i\Delta t}(\Omega_{A}e^{i\phi}|A\rangle\langle p|+{\Omega_{B}}|B\rangle\langle p|)\cr\cr&&
	+ \frac{1}{2}e^{i(\Delta+V)t}{\Omega_{C}}|R\rangle\langle p|+ {\rm H.c.}
	\end{eqnarray}
The effective Hamiltonian of Eq.~(\ref{e10j}) can be rewritten as~\cite{dj2007}
	\begin{eqnarray}\label{e11j}
	\hat{H}_{\rm s}^{\rm II} = \frac{\Omega'^{2}}{4\Delta}|\mathcal{B}\rangle\langle \mathcal{B}|
	\end{eqnarray}
where we have discarded the stark shifts relevant to $|p\rangle\langle p|$ and $|R\rangle\langle R|$, which have no influence on the system because initial state is in the ground state subspace and these two terms have no energy exchange with the ground state in the whole evolution process. After considering the operations of canceling stark shifts in case I~[the same as operations from Eq.~(\ref{e03n}) to Eq.~(\ref{e04n})], Eq.~(\ref{e11j}) is vanished. That is, for case II, each of the ensemble atom would keep invariant, which means the ensemble qubit keeps invariant. It should be noted that, although the stark shifts of $|R\rangle\langle R|$ for cases I and II are different, the operations we perform to cancel the stark shifts are same. And whether the stark shift of $|R\rangle\langle R|$ is canceled out or not for case II has no influence on the scheme since the initial state is in ground state subspace and the Rydberg states are decoupled with the ground state subspace for case II. Summarizing the discussion above, we obtain the evolution operator of step~(ii)
	\begin{equation}
	\hat{U}_{\rm ii} =|0\rangle_{c}\langle0|\otimes\hat{\mathcal{U}}+ |r\rangle_{c}\langle r|\otimes\hat{\mathcal{I}}
	\end{equation}
	\begin{figure}
	\includegraphics[width=1\linewidth]{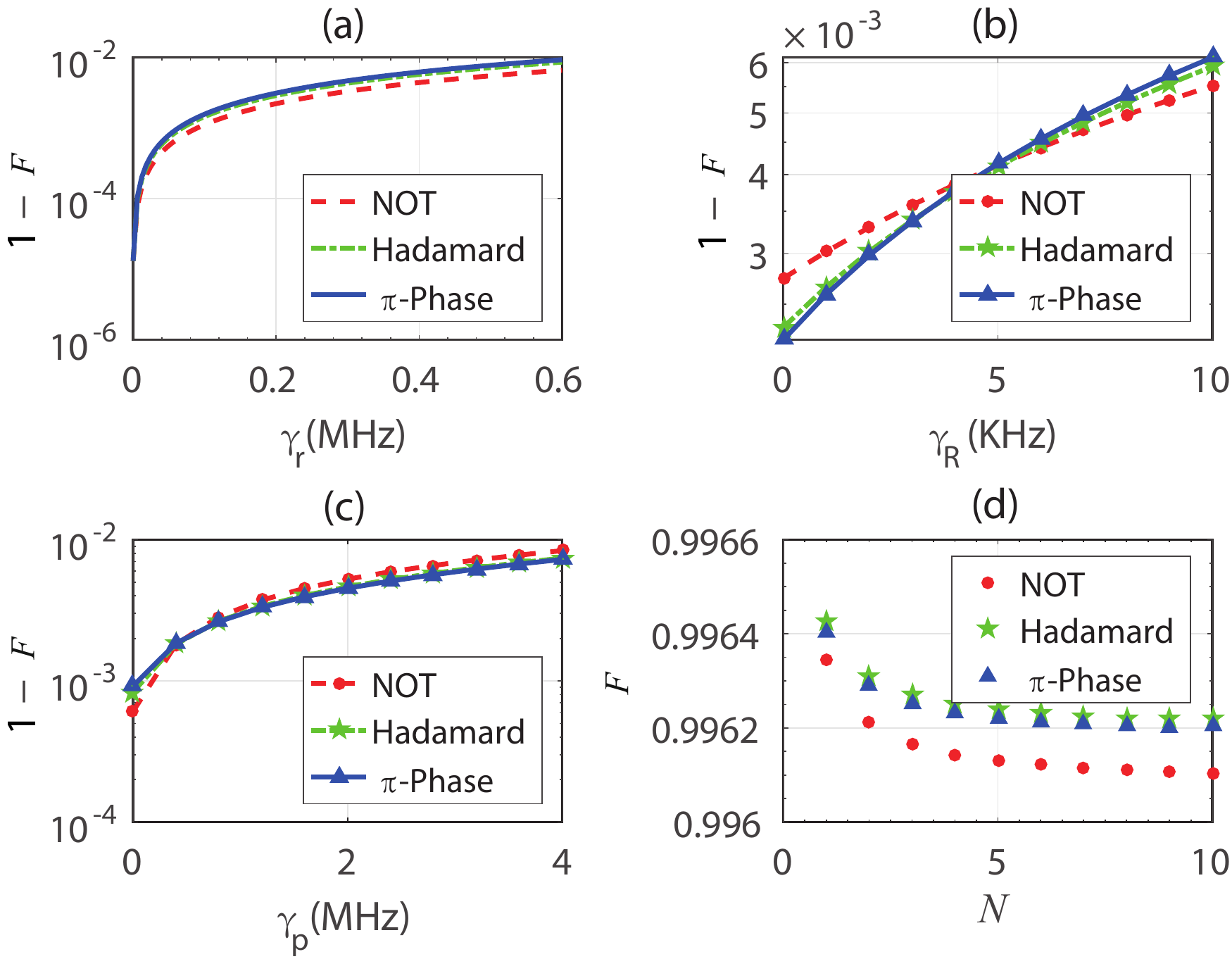}\\
	\caption{(a) Average fidelity of NHQC gates for control-qubit versus atomic spontaneous emission rate of $|r\rangle$ state. The parameters are  $\Omega_{1} = 2\pi\times10$~MHz, $\Omega_{0}=\Omega_{1}\tan{(\theta/2)}$, and the final evolution time $T$ of panel (a) is determined by the condition $\int_{0}^{T}\Omega dt=2\pi$. In (b) and (c), we show the average fidelity of the NHQC gate carried out by the target ensemble qubit versus decay rate in $|R\rangle$ and $|p\rangle$ states, respectively. The ensemble atom number is $N=4$. The parameters are chosen as $\Omega_{B} = 2\pi\times10$~MHz, $\Omega_{A}=\Omega_{B}\tan(\vartheta/2)$, $\Omega_{C} =\Omega_{B}$, $\Delta=12\Omega_{B}$, $\gamma_{p}=1$~MHz [in (b)], and ~$\gamma_{R}=4$~kHz [in (c)].
	(d) Average fidelity versus the number of ensemble atom \emph{N}. The parameters are $\Omega_{B}=2\pi\times10$~MHz, $\Omega_{A}=\Omega_{B}\tan(\vartheta/2)$, $\Omega_{C} =\Omega_{B}$, $\Delta=12\Omega_{B}$, $\gamma_{p}=1$~MHz and $\gamma_{R}=4$~kHz. In panels (b), (c) and (d), the final evolution time $T$ is determined by the condition ${\int_{0}^{T}\Omega'\Omega_{c}/(2\Delta) dt = 2\pi}$.}\label{f003}
	\end{figure}

Step~(iii): Perform the inverse operation of step~(i) by setting $\varphi_{1}=\pi$. In this case, the Hamiltonian is given as
\begin{equation}\label{e14}
\hat{H}_{c} =\frac{\Omega_{1}e^{i\pi}}{2}|1\rangle\langle r|+ {\rm H.c.}.
\end{equation}
If condition $\int_{0}^{\tau_{c}/2}\Omega_{1}(t)dt=\pi$ is satisfied, $|r\rangle\rightarrow i|1\rangle$ is achieved. After the whole steps, the two-qubit gate is described by the evolution operator,
\begin{equation}
\hat{U}_{\rm two}=|0\rangle_{c}\langle0|\otimes\hat{\mathcal{U}}+ |1\rangle_{c}\langle 1|\otimes\hat{\mathcal{I}}.
	\end{equation}
	\begin{figure}
	\includegraphics[width=\linewidth]{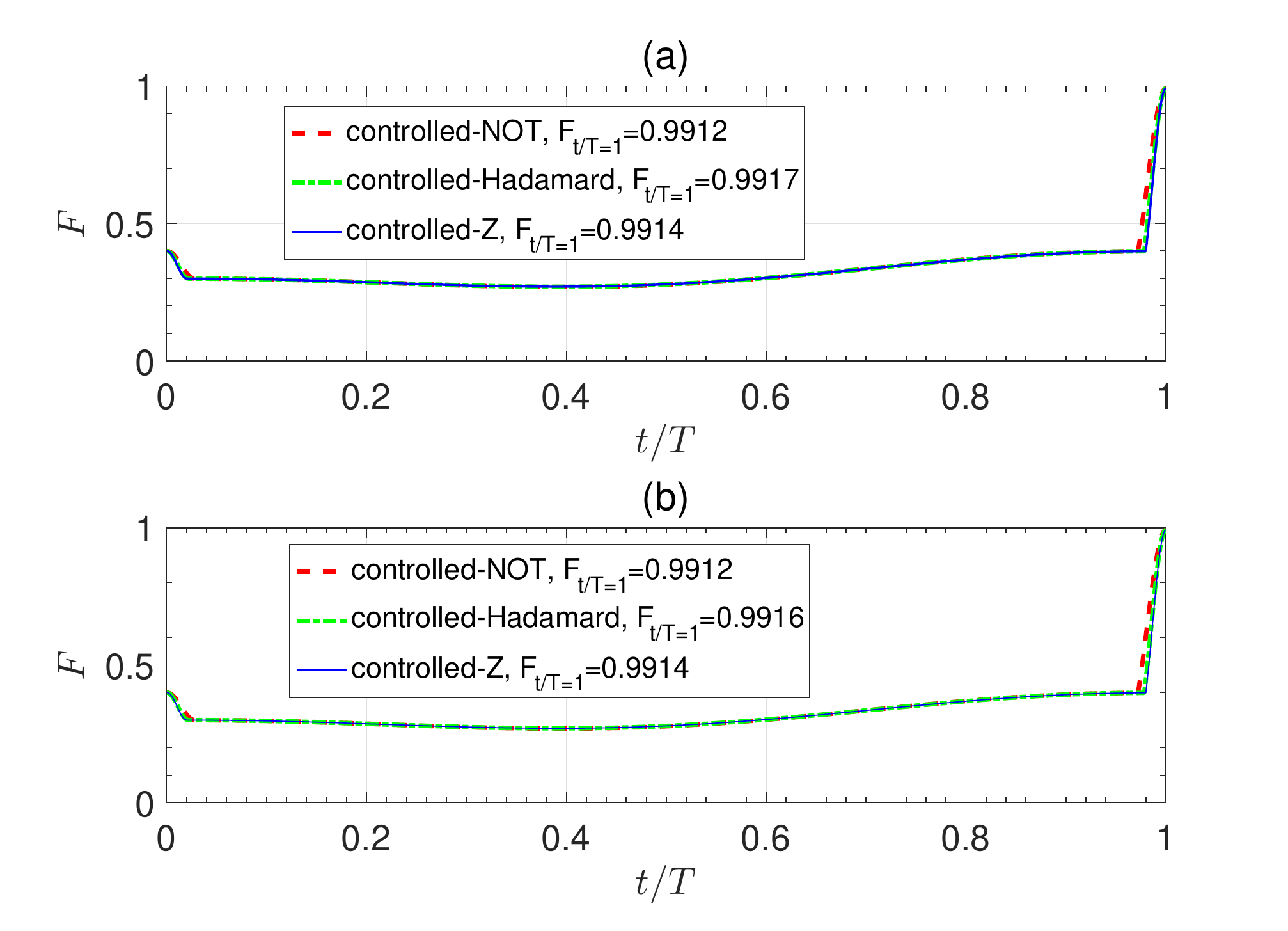}\\
	\caption{Evolution of the average fidelity of two-qubit NHQC gates when numbers of atoms in the ensemble are $N=4$ (a) and 8 (b). The parameters are chosen as $\Omega_{0}=0$, $\Omega_{1}=2\pi\times10$~MHz, $\Omega_{B}=\Omega_{1}$, $\Omega_{A}=\Omega_{B}\tan(\vartheta/2)$, $\Omega_{C}=\Omega_{B}$, $\Delta=12\Omega_{B}$, $V=2\Delta$. $\gamma_{r}=\gamma_{R}=4$~kHz and $\gamma_{p}=1$~MHz.}\label{f004}
	\end{figure}

Now lets check the holonomy of the scheme. For target ensemble, the holonomy has been discussed in step~(ii). For control atom, we can specify $|d\rangle$ and $|b\rangle$ as $|0\rangle$ and $|1\rangle$, respectively. After considering steps (i) and (iii), $|b\rangle\rightarrow-i|r\rangle\rightarrow|b\rangle$ is realized and $|d\rangle$ is always invariant, which means the cyclic condition is satisfied as well. Also, the parallel-transport condition is satisfied: $\langle\psi_k{(t)}| \hat{H}_{c}|\psi_l{(t)}\rangle=\langle \psi_k{(0)}| U^\dagger \hat{H}_{c}U|\psi_l{(0)}\rangle=0$ with $\{|\psi_k{(0)}\rangle,|\psi_l{(0)}\rangle\}\in\{|0\rangle,|1\rangle\}$.

\subsection{Gate fidelities}\label{s4}
The performance of different gates will be affected by dissipation processes, even when laser parameters are ideal. To take into account these processes, we solve the master equation numerically and evaluate gate fidelities. In Fig.~\ref{f003}, we plot the average fidelity of single-qubit and single-ensemble-qubit NHQC gates with respect to atomic spontaneous emission rate. One can see in Fig.~\ref{f003}(a) that the fidelity can be as high as 0.999 if $\gamma_{r}$ is less than 0.06~MHz. This is achievable as typical Rydberg lifetimes range from $10\,\mu$s to $100\,\mu$s~\cite{beterov_quasiclassical_2009}. In Fig.~\ref{f003}(b) and (c), the average fidelity of the logic qubit made of the Rydberg atom ensemble is shown. By varying the atomic spontaneous emission in the Rydberg state (b) or the intermediate state (c), the fidelity is around 0.999. Note that in the simulation, we have used the full Hamiltonian but not the effective Hamiltonian. For an ideal gates obtained from effective Hamiltonian, the higher order terms that were neglected will cause gate errors. Fig.~\ref{f003}(d) shows the average fidelity of ensemble-qubit versus the number of ensemble atom \emph{N}. One can see that the gate fidelity is weakly depending on \emph{N}. Increasing the number of atoms, the fidelity decreases negligibly.

In Fig.~\ref{f004}, dynamical evolution of the average fidelity for two-qubit NHQC gates is shown when the ensemble atom number $N=4$ and 8. We have used different sets of parameters, i.e. $\{\vartheta, \phi\}$ equals $\{-\pi/2, 0\}$ for controlled-NOT gate, $\{-\pi/4, 0\}$ for controlled-Hadamard gate, and $\{0, 0\}$ for controlled-Z gate, respectively. The fidelity for all the gates are above 0.99 for both $N=4$ and $8$.
Moreover, we simulate dynamical evolution of two-qubit controlled-NOT gate, where the fidelity can reach 0.9971 without dissipation for $N=4$ and $N=8$. These results demonstrate that the scheme is robust and insensitive to the number of atoms in the ensemble when $N\ge2$.

\section{Optimized Geometric gates}\label{s20}

We will first introduce the NHQC+ scheme and the respective requirements in the dynamics. To implement the scheme using the Rydberg interaction, we construct the single-ensemble-qubit gate via dynamical-invariant-based inverse engineering in Sec.~\ref{s4.1}. Based on the process of Sec.~\ref{s4.1}, we further use the optimal method to show that the gate is robust even when certain systematic errors are present in the dynamics in Sec.~\ref{slogicalqubit}. The optimized two-qubit case is illustrated in Sec.~\ref{s3.3}. For single control atom, the optimized method is similar to the single-ensemble-qubit case, and we will not consider it here.
\subsection{Requirements of the NHQC+ scheme}

To combine NHQC with the optimal control theory, we here consider to break the parallel-transport condition of NHQC by following the method in Refs.~\cite{lsx2019,ylx2019}, i.e., NHQC+ dynamics, by the inverse engineering. The dynamical phase is canceled out entirely by dividing the whole evolution process into two parts with opposite dynamical effect, which is not the case of NHQC where the dynamical effect is zero at any time through satisfying the parallel-transport condition.

For a general time-dependent Hamiltonian $\hat{H}(t)$, we consider one complete set of basis vector, $\{|\Psi_{m}(0)\rangle\}$ at $t=0$. And the evolution of the time-dependent state $|\Psi_{m}(t)\rangle$ follows the Schr{\"o}dinger equation. We now choose another set of basis $|\nu_{m}(t)\rangle$, which is connected to $|\Psi_{m}(t)\rangle$ through unitary transformation. The following three conditions should be satisfied for the NHQC+ dynamics~\cite{lsx2019}. i) cyclic condition, i.e. $|\nu_{m}(0)\rangle=|\Psi_{m}(0)\rangle$ and $|\nu_{m}(\tau)\rangle$ should both evolve back to initial state $|\nu_{m}(0)\rangle$. $t=0$ and $t=\tau$ denote initial and final moment, respectively. ii) $|\nu_{m}(t)\rangle\langle\nu_{m}(t)|$ should satisfy von Neumann equation $\frac{d}{dt}|\nu_{m}(t)\rangle\langle\nu_{m}(t)|=-i[\hat{H},~|\nu_{m}(t)\rangle\langle\nu_{m}(t)|]$. Finally iii) the dynamical phase vanishes at the end of the evolution, $\int_{0}^{\tau}\langle\nu_{k}(t)|\hat{H}(t)|\nu_{k}(t)\rangle dt=0$.

\subsection{Invariant-based inverse engineering of single-ensemble-qubit gate}\label{s4.1}
\subsubsection{Theoretical analysis}
We first rewrite Hamiltonian~(\ref{e07n}) in a matrix form,
	\begin{eqnarray}\label{e20j}
	\hat{\mathcal{H}}_{\rm eff}=\frac{1}{2}\left(
	\begin{array}{cccc}
	0 & \Omega_{R}-i\Omega_{I} \\
	\Omega_{R}+i\Omega_{I} & 0 \\
	\end{array}
	\right),
	\end{eqnarray}
where  $\Omega_{R}=\Omega_{\rm eff}\cos\phi_{B}$ and $\Omega_{I}=-\Omega_{\rm eff}\sin\phi_{B}$ with $\Omega_{\rm eff} = \Omega'\Omega_{C}/(2\Delta)$.

Here we make use of the Lewis-Riesenfeld invariants~\cite{LRinvariant,Muga_2009,chen2010,Ruschhaupt_2012,del2011} to implement the quantum gate. In this approach, the Hermitian operator $\hat{I}(t)$ of a dynamical invariant satisfies $\partial\hat{I}(t)/\partial t+i[\hat{\mathcal {H}}_{\rm eff},\hat{I}]=0$. Knowing the Hamiltonian $\hat{\mathcal {H}}_{\rm eff}$, we can obtain the expression of $\hat{I}(t)$ explicitly as
	\begin{eqnarray}\label{e21j}
	\hat{I}(t)=\frac{\mu}{2}\left(
	\begin{array}{cccc}
	\cos[\Theta(t)] &  e^{-i\alpha(t)}\sin[\Theta(t)] \\
	e^{i\alpha(t)}\sin[\Theta(t)] & -\cos[\Theta(t)] \\
	\end{array}
	\right),
	\end{eqnarray}
	where $\mu$ is an arbitrary constant, $\dot{\Theta}(t)=\Omega_{I}\cos\alpha-\Omega_{R}\sin\alpha$, $\dot{\alpha}=-\cot\Theta(\cos\alpha\Omega_{R}+\sin\alpha\Omega_{I})$. The orthogonal eigenvector of the invariant $\hat{I}(t)$ with the eigenvalues $\pm\mu/2$ reads
	\begin{eqnarray}\label{e22j}
	|\phi_{+}(t)\rangle  =  \cos(\Theta/2)e^{-i\alpha/2}|\overline{\mathcal{B}}\rangle + \sin(\Theta/2)e^{i\alpha/2} |\overline{R}\rangle,\cr\cr
	|\phi_{-}(t)\rangle   =    \sin(\Theta/2)e^{-i\alpha/2}|\overline{\mathcal{B}}\rangle - \cos(\Theta/2)e^{i\alpha/2}|\overline{R}\rangle.
	\end{eqnarray}
	Then the wavefunction $|\Psi_{m}(t)\rangle$ which follows the Schr{\"o}dinger equation can be generally written as $|\Psi_{m}(t)\rangle=c_{+}(t)e^{if_{+}(t)}|\phi_{+}(t)\rangle+c_{-}(t)e^{if_{-}(t)}|\phi_{-}(t)\rangle$. Here $c_{\pm}$ is complex constant coefficient and $\dot{f_{\pm}}=\langle\phi_{\pm}|i\frac{\partial}{\partial t}-\hat{\mathcal{H}}_{\rm eff}|\phi_{\pm}(t)\rangle$ denotes the Lewis-Riesenfeld phase. We choose the orthogonal solution of the Schr{\"o}dinger equation as
	\begin{eqnarray}\label{e23}
	&&|\psi(t)\rangle= |\phi_{+}(t)\rangle e^{-i\gamma(t)/2}\cr\cr
	&&|\psi_{\bot}(t)\rangle= |\phi_{-}(t)\rangle e^{i\gamma(t)/2}.
	\end{eqnarray} Obviously, $\gamma=-2f_{+}=2f_{-}$ and $\dot{\gamma}=(\cos\alpha\Omega_{R}+\sin\alpha\Omega_{I})/\sin\Theta$ can be achieved through the Lewis-Riesenfeld phase. On the other hand, suppose the solution of Schr{\"o}dinger equation as $|\psi(t)\rangle$~($|\psi(t)\rangle\langle\psi(t)|$ is a dynamical invariant) and substitute it into the Schr{\"o}dinger equation, one can also get dynamical equations of $\dot{\Theta},~\dot{\alpha}$, and $\dot{\gamma}$, respectively. Shapes of the pulse $\Omega_{R}$ and $\Omega_{I}$ can be obtained through using $\dot{\Theta},~\dot{\alpha}$, and $\dot{\gamma}$~[see Appendix~\ref{s8} and~\ref{appc1} for details].

  To study the NHQC+ dynamics, we choose the auxiliary basis as $|\nu_{0}\rangle=|\overline{\mathcal{D}}\rangle$ and $|\nu_{1}(t)\rangle=|\phi_{+}(t)\rangle$. Since $|\overline{\mathcal{D}}\rangle$ is the dark state of the system, we here only check whether $|\nu_{1}(t)\rangle$ satisfies the conditions of NHQC+. Here we assume $\Theta(t) = 2\pi t/\tau$, $\gamma(t) = 2\Theta$ and $\alpha(t)=-2\sin\Theta$.
  i) For the cyclic condition, since $\Theta(0)=0$, $\Theta(\tau)=2\pi$, $\alpha(0)=0$, $\alpha(\tau)=0$, and $\gamma(0)=0$, one can get $|\psi(0)\rangle=|\nu_{1}(0)\rangle=-|\nu_{1}(\tau)\rangle$, which means the cyclic condition is satisfied. ii) Since $|\psi(t)\rangle$ follows the Schr{\"o}dinger equation, it can be easily verified that $|\nu_{1}(t)\rangle$ satisfies von Neumann equation~[see Appendix ~\ref{s9}]. iii) Then $\Omega_{R}$ suddenly have a minus sign at the half moment of the evolution for optimized NHQC gates. One can check that $\int_{0}^{\tau/2}\langle\nu_{1}(t)|\hat{\mathcal{H}}_{\rm eff}|\nu_{1}(t)\rangle dt=-\int_{\tau/2}^{\tau}\langle\nu_{1}(t)|\hat{\mathcal{H}}_{\rm eff}|\nu_{1}(t)\rangle dt$ [see Appendix ~\ref{s10}], which means the third condition of NHQC+ is satisfied. Thus, all of the conditions of NHQC+ dynamics are met.

	\begin{figure}
		\includegraphics[width=\linewidth]{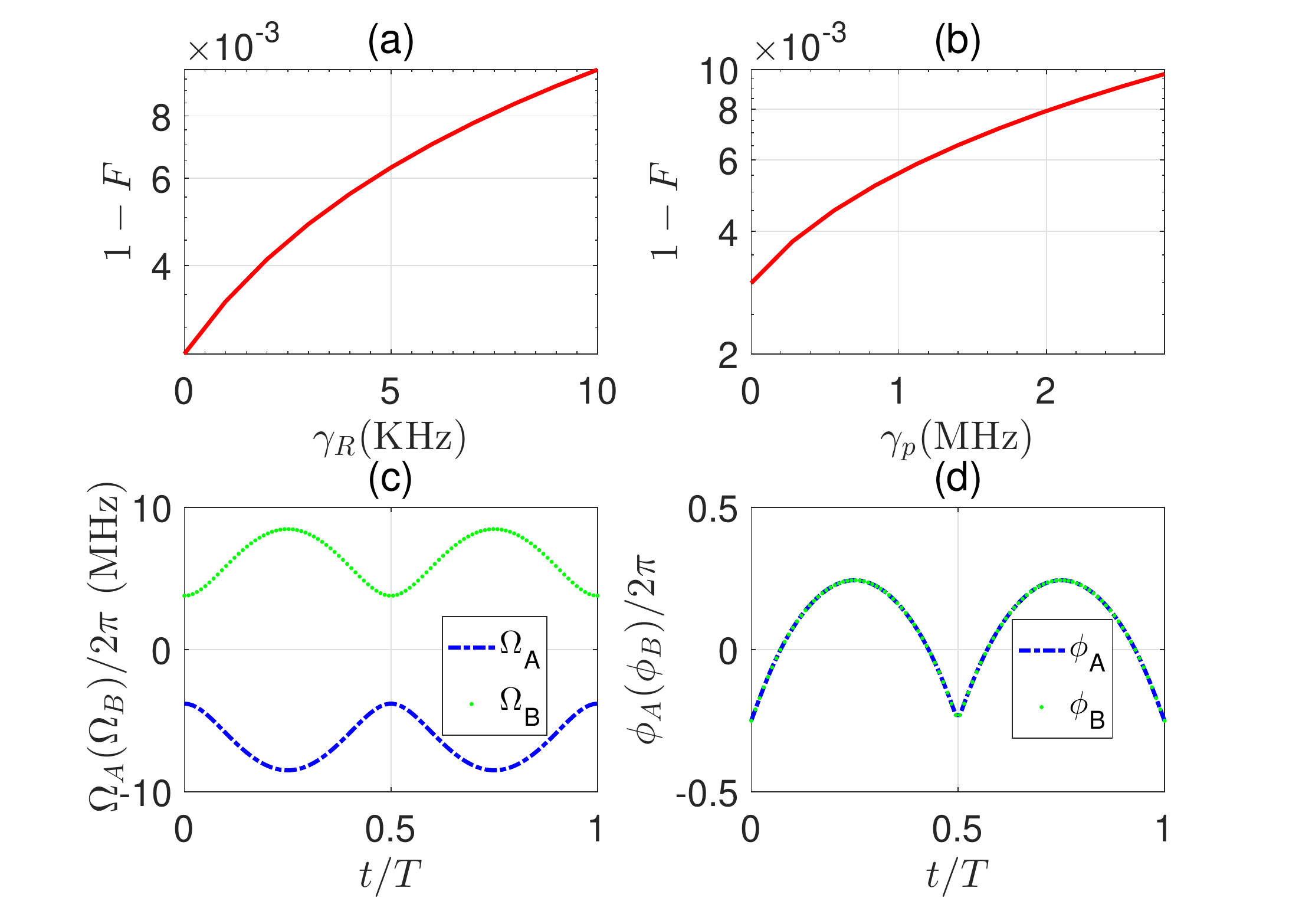}\\
		\caption{Average fidelity of the NOT gate obtained using the invariant-based method. (a)[(b)] Average fidelity versus $\gamma_{R}$($\gamma_{p}$). (c) and (d) show the pulse shape and phase information. max$(\Omega_{\rm eff})/2\pi=0.5$~MHz is set in advance. Other parameters are chosen as $\Omega_{C}/2\pi=10$~MHz, $\Delta=12\Omega_{C}$, $\Theta(t) = 2\pi t/\tau$, $\gamma(t) = 4\pi t/\tau$, $\alpha(t)=-2\sin\Theta$. $\{\vartheta, \phi\}$ equals $\{-\pi/2, 0\}$. 
		$\gamma_{p}=1$~MHz and $\gamma_{R}=4$~kHz~[For (a) and (b), we keep one of these two rates fixed, and change the other]. The number of ensemble atoms is $N=4$.}\label{f005}
	\end{figure}
As an example,  we plot the average fidelity of the ensemble-qubit NOT gate and pulse profiles, respectively in Fig.~\ref{f005}. The reason why the fidelity is slightly lower than that of the same gate in Fig.~\ref{f003} is that the Rabi frequencies $\Omega_{A(B)}$ we employed here is small, which gives a gate time more than twice that of Fig.~\ref{f003}. Thus, the influence of dissipation increases. In the following subsection, we will demonstrate the robustness of the invariant-based optimal scheme with respect to parameter fluctuations.

\subsection{Optimized single ensemble qubit gate with ZSS optimal control}\label{slogicalqubit}
We will first optimize the performance of the scheme when there are static errors in the parameters. To be concrete, we consider that $\Omega_{A}$, $\Omega_{B}$, $\Omega_{C}$ and $\Delta$ may have some fluctuations as seen in typical experiments. In our analysis, we assume $\Omega_{\rm eff}$ becomes $\Omega_{\rm eff}\rightarrow(1+\varepsilon)\Omega_{\rm eff}$ where $\varepsilon$ is a small influence representing a systematic error. With this parameter fluctuation, Hamiltonian~(\ref{e07n}) becomes
\begin{equation}\label{e024}
\hat{\mathcal{H}}_{\rm eff}'
=\frac{e^{i\phi_{B}}(1+\varepsilon)\Omega_{\rm eff}}{2}|\overline{\mathcal{B}}\rangle\langle \overline{R}|+ {\rm H.c.},
\end{equation}
with $\Omega_{\rm eff} = \Omega'\Omega_{C}/(2\Delta)$.

We then apply the ZSS~\cite{Ruschhaupt_2012} optimal protocol, in which the systematic-error sensitivity is defined as $q_{s}=-\frac{1}{2}\frac{\partial^{2}P}{\partial\varepsilon^{2}}|_{\varepsilon=0}=\left|\int_{0}^{\tau/2}dt\langle\psi_{\bot}(t)|\hat{\mathcal{H}}_{\rm eff}|\psi_{0}(t)\rangle\right|^{2}$. And \emph{P} denotes the probability to be excited to $|\overline{R}\rangle$ at the half evolution time $\tau/2$ in our scheme. Combining Eqs.~(\ref{e20j}),~(\ref{e23}) and (\ref{e25j}), one can obtain the expression of $q_{s}=\left|\int_{0}^{\tau/2}dte^{-i\gamma}\dot{\Theta}\sin^{2}\Theta\right|^{2}$~[see Appendix ~\ref{s11}].

\begin{figure}
    \centering
    \includegraphics[width=\linewidth]{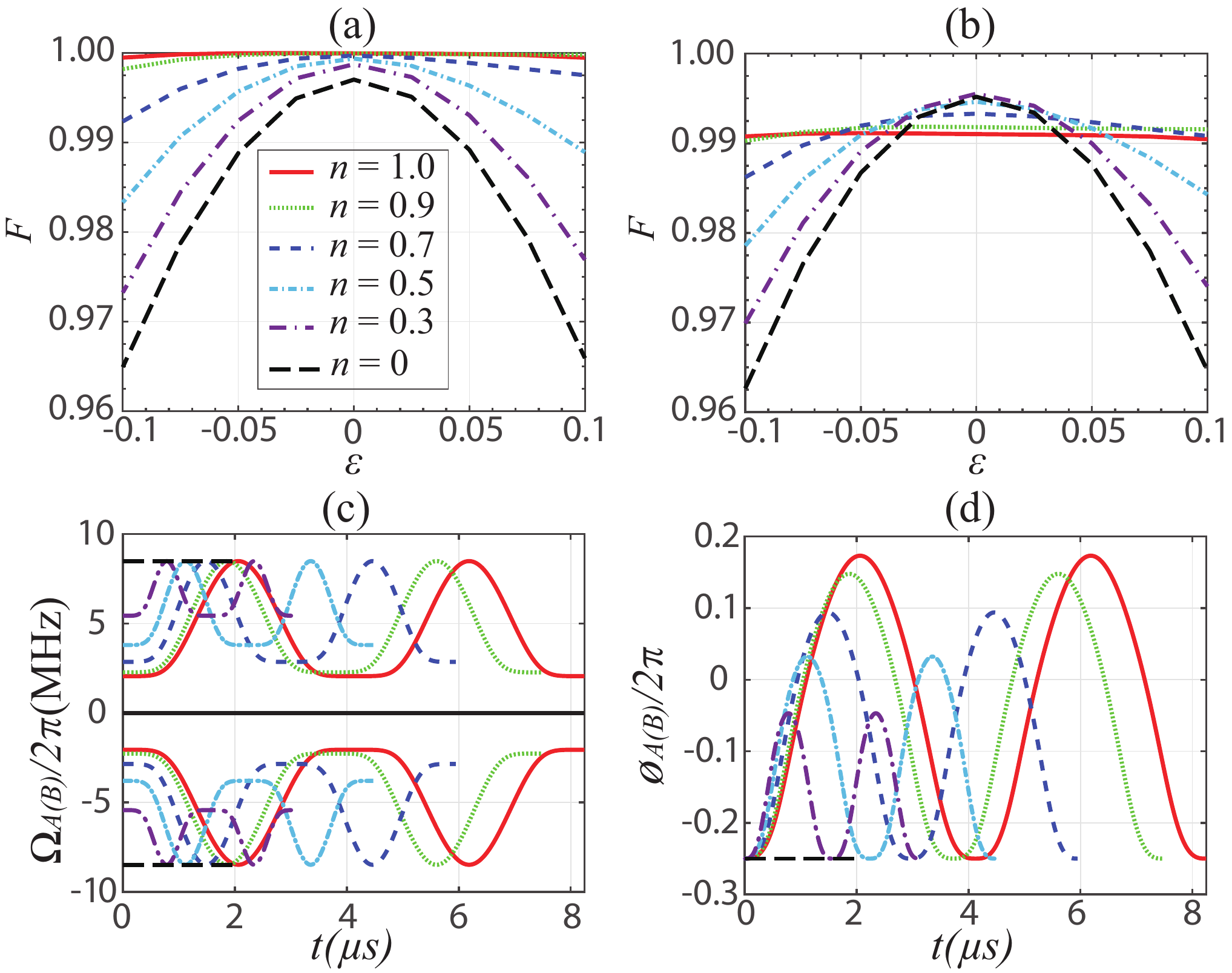}
    \caption{Average fidelity of single-ensemble-qubit NOT gate versus systematic error $\varepsilon$ with different optimized parameters $n$ without (a) and partially considering (b) dissipation $\gamma_{R}=4$~kHz and $\gamma_{p}=1$~MHz. The legend in (b) is the same as (a). The rest parameters are $\Omega_{C}/2\pi=10$~MHz, $\Delta=12\Omega_{C}$, $\Theta(t) = 2\pi t/\tau$, $\gamma(t) = n[2\Theta-\sin(2\Theta)]$, $\alpha(t) = -4n\sin^3\Theta/3$. $\{\vartheta, \phi\}$ equals $\{-\pi/2, 0\}$. $\Omega_{A}$, $\Omega_{B}$, $\phi_{A}$ and $\phi_{B}$ are calculated based on Eq.~(\ref{e26j}). Panels (c) and (d) show Rabi frequencies and phases shape for different optimized parameter \emph{n}, respectively. The maximal value $\Omega_{\rm eff}(t)/2\pi$ is $0.5~{\rm MHz}$. In panel (d) $\phi_{A}$ and $\phi_{B}$ are same.}
    \label{f006}
\end{figure}

To minimise $q_{s}$, we first set $\gamma(t)=n[2\Theta-\sin(2\Theta)]$, which further leads to $q_{s}=\sin^{2}(n\pi)/4n^{2}$. It is easy to show when $n\rightarrow0$, $q_{s}\rightarrow\pi^{2}/{4}$, which recovers the previous NHQC scheme. When $n$ is the positive integer (i.e., $n=1,2,3,...$), $q_{s}=0$, so we achieve the minimum value of $q_{s}$. In Fig.~\ref{f006}(a), we can see that $n=1$ gives the most robust situation without considering dissipation. For simplicity, here we choose $\Theta(t) = 2\pi t/\tau$, $\alpha(t) = -4n\sin^3\Theta/3$.

In Fig.~\ref{f006}, we plot the average fidelity of single-ensemble-qubit NOT gate with different optimized parameters versus systematic errors when dissipation is fully or partially turned off. The systematic error varies from $-0.1$ to $0.1$, and we choose the value of $n$  between $0$ and $1$. As shown in Fig.~\ref{f006}(a), the  average fidelity is improved by increasing of $n$. When partially considering dissipation in Fig.~\ref{f006}(b), the fidelity is no longer a monotonic function of $n$. Roughly, the fidelity increases with the increase of $n$ when $0.05<\left|\varepsilon\right|<0.1$. And the fidelity is negatively correlated with $n$~($n>0$) when $\left|\varepsilon\right|$ is between $0$ and $0.02$. That is because greater $n$ corresponding to longer evolution time, where the dissipation plays more important roles. Therefore, with the consideration of dissipation, one should choose the optimized parameter \emph{n} carefully based on the trend of average fidelity with respect to $\varepsilon$ for concrete systems [similar to Fig.~\ref{f006}(b)]. Fig.~\ref{f006}(c) and (d) show the optimal Rabi frequencies and phases of the designed pulses, respectively, for different optimized parameter \emph{n}.

We now discuss the case when $\varepsilon=0$ in Fig.~\ref{f006}(a). One can see that the fidelity when $n=0$ (i.e., the conventional NHQC gates) is less than the fidelity when $n\neq0$ and the fidelity increases as \emph{n} increases, which means that the optimized scheme has advantages even without systematic error. This phenomena can be understood from Fig.~\ref{f006}(c), the mean values of $\Omega_{A}$ and $\Omega_{B}$ decreases as \emph{n} increases, which means that the large detuning condition $\Delta\gg\Omega'$ would be better satisfied as \emph{n} increases. Thus, the error induced by ignoring high-order perturbation terms is decreased and the fidelity increases.

\subsection{Optimized two-qubit gate}\label{s3.3}
To consider the optimized two-qubit quantum logic gate regarding to the systematic error, we first rewrite the Hamiltonian of single control atom in Eq.~(\ref{e04}) as
	\begin{equation}\label{e024n}
\hat{H}_{c} =\frac{\Omega_{1}e^{i\varphi_{1}}}{2}|1\rangle\langle r|+ {\rm H.c.},
\end{equation}
which has similar form as Eq.~(\ref{e07n}). Thus, one can use the method similar as that in Sec.~\ref{slogicalqubit} to design pulses of control atom to achieve the desired process. The difference is that in the middle of the evolution of the control atom (the time when the control atom is excited), the laser needs to be turned off, and at the same time the laser of the target atom is turned on. The other half of the control atom's pulse needs to turn on until the operation of the target atom is completed. For the target atom, the parameters are the same as that in Sec.~\ref{slogicalqubit}. Concretely, pulses of the two-qubit gate are
\begin{equation} \label{e025n}
\left\{ \begin{aligned}
\Theta_{c}& = 2\pi t/\tau_{c},~~(0<t\leqslant\tau_{c}/2)\\
\Theta_{c}& = 2\pi(t-\tau_{t})/\tau_{c},~~ (\tau_{c}/2+\tau_{t}<t\leqslant\tau_{c}+\tau_{t}) \\
\Theta_{t}& = 2\pi(t-\tau_{c}/2)/\tau_{t},~~(\tau_{c}/2<t\leqslant\tau_{t}+\tau_{c}/2)\\
\end{aligned} \right.,
\end{equation}
in which footnotes \emph{c} and \emph{t} denote control atom and target ensemble atom, respectively. Other parameters can be obtained by using Eq.~(\ref{e26j}).

For target ensemble atom, all of the chosen parameters are the same as that in Sec.~\ref{slogicalqubit}, which means that the integral of dynamical phase is zero. For the control atom,  one can also check that the integration in Eq.~(\ref{e33j}) is zero with the expressions in Eq.~(\ref{e025n}), which means the dynamical phase in the whole process is zero. This shows that the evolution of the two-qubit gate satisfies the requirement of the NHQC+ scheme.

\begin{figure}
	\centering
	\includegraphics[width=\linewidth]{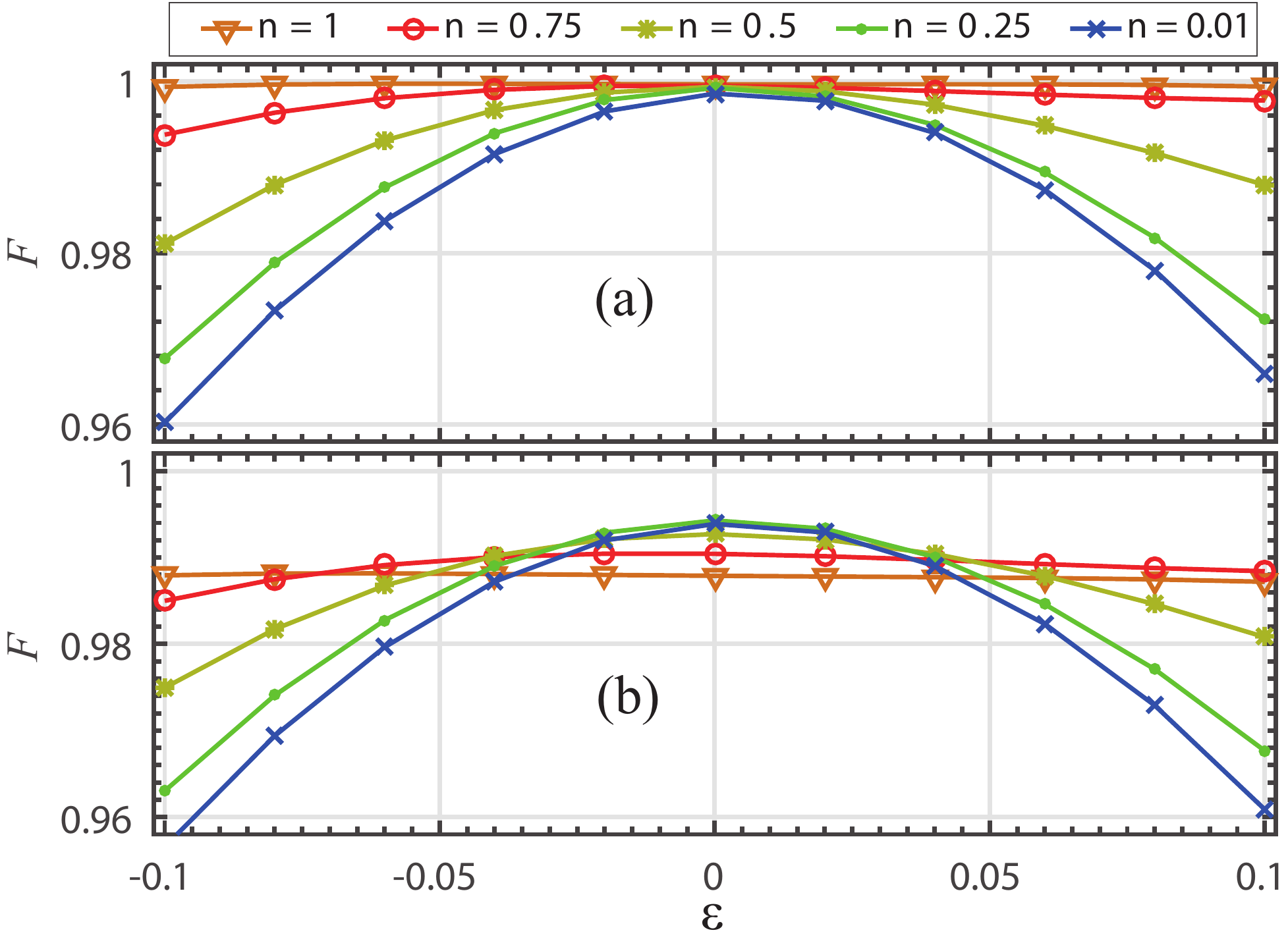}
	\caption{Average fidelity of two-qubit controlled-NOT gate versus systematic errors $\varepsilon$ with different optimized parameters $n$ (a) without considering dissipation and (b) with $\gamma_{r}=\gamma_{R}=2$~kHz and $\gamma_{p}=2$~MHz. $\varepsilon_{c}=\varepsilon_{t}=\varepsilon$ and $n_{c}=n_{t}=n$ are set for simplicity, where footnote \emph{c} and \emph{t} denote control and target ensemble atom, respectively. The parameters are chosen as $\Omega_{C}/2\pi=10$~MHz, $\Delta=12\Omega_{C}$, $V=0.9\Delta$, $\{\vartheta, \phi\}$ for target atom equals $\{-\pi/2, 0\}$. We choose ${\rm Max}[\Omega_{1}(t)/2\pi] = 6~{\rm MHz}$ and ${\rm max}[\Omega_{\rm eff}(t)/2\pi] = 0.5~{\rm MHz}$. Other parameters are calculated based on Eqs.~(\ref{e025n}) and ~(\ref{e26j}), respectively.}
	\label{f007}
\end{figure}

In Fig.~\ref{f007}, we plot the average fidelity of two-qubit controlled-NOT gate versus the static systematic error $\varepsilon$ with different optimized parameters. One can see that, without considering dissipation [see Fig.~\ref{f007}(a)], the performance of the scheme becomes better as the value of optimized parameter $n$ increases. In Fig.~\ref{f007}(b) we show the average fidelity when the dissipation is considered. The fidelity is slightly reduced because the larger \emph{n} corresponds to the longer evolution time. The dissipation thus impacts the gate fidelity stronger. For concrete experiments, one would like to achieve higher fidelities and shorter gate times. According to Fig.~\ref{f007}(b), we can choose, for example, $n=0.5$ or $n=0.75$ to achieve this goal.

	\section{DISCUSSIONS}\label{s5}

	\subsection{Theory comparison}
	In comparison with Ref.~\cite{mihhp2009} which inspires us the basic model, our scheme mainly has the following differences: i) The RRI strength among the ensemble atom of our scheme would have no influence on the performance once the ensemble qubit is prepared. That is because only one atom is in Rydberg state for our ensemble qubit. While in Ref.~\cite{mihhp2009}, the RRI among the ensemble atoms should be less than 0.4$\epsilon$ (here $\epsilon$ is defined as characteristic energy scale in Ref.~\cite{mihhp2009}) to ensure the high fidelity. ii) More universal controlled gates rather than controlled-NOT gate can be constructed through modulating laser parameters. iii) $\Omega_{C}$ has the same order of magnitude with  $\Omega_{A}$ and $\Omega_{B}$. These differences may relax the experimental requirements and broaden the application range. Recently, through introducing photon freedom assisted by the microwave field and considering NHQC pulse, Ref.~\cite{zhaoensemble} demonstrated numerically two-qubit swap gate with the fidelity about is close to 0.83 including dissipation, which can be improved close to 0.95 after considering broad laser parameters. In contrast to Ref.~\cite{zhaoensemble}, the NHQC scheme here has higher fidelities even without optimized pulse. In contrast to Ref.~\cite{kang2018} which construct NHQC gates via shortcut-to-adiabaticity between two single atoms, the basic dynamical process is different. And our scheme focuses on the ensemble qubit and also studies how to further enhance the robustness via the optimal control method.

   \subsection{More general cases}
         In this subsection, we consider our schemes with more general or practical cases, including the multiple-qubit case, compatibility to Rydberg dark state dynamics, exchanging the roles between single atom and ensemble. For simplicity, here we only consider the NHQC dynamics. And the NHQC+ dynamics is also feasible for these schemes if we add more controls on the Hamiltonian.
   \begin{figure}
   	\centering
   	\includegraphics[width=\linewidth]{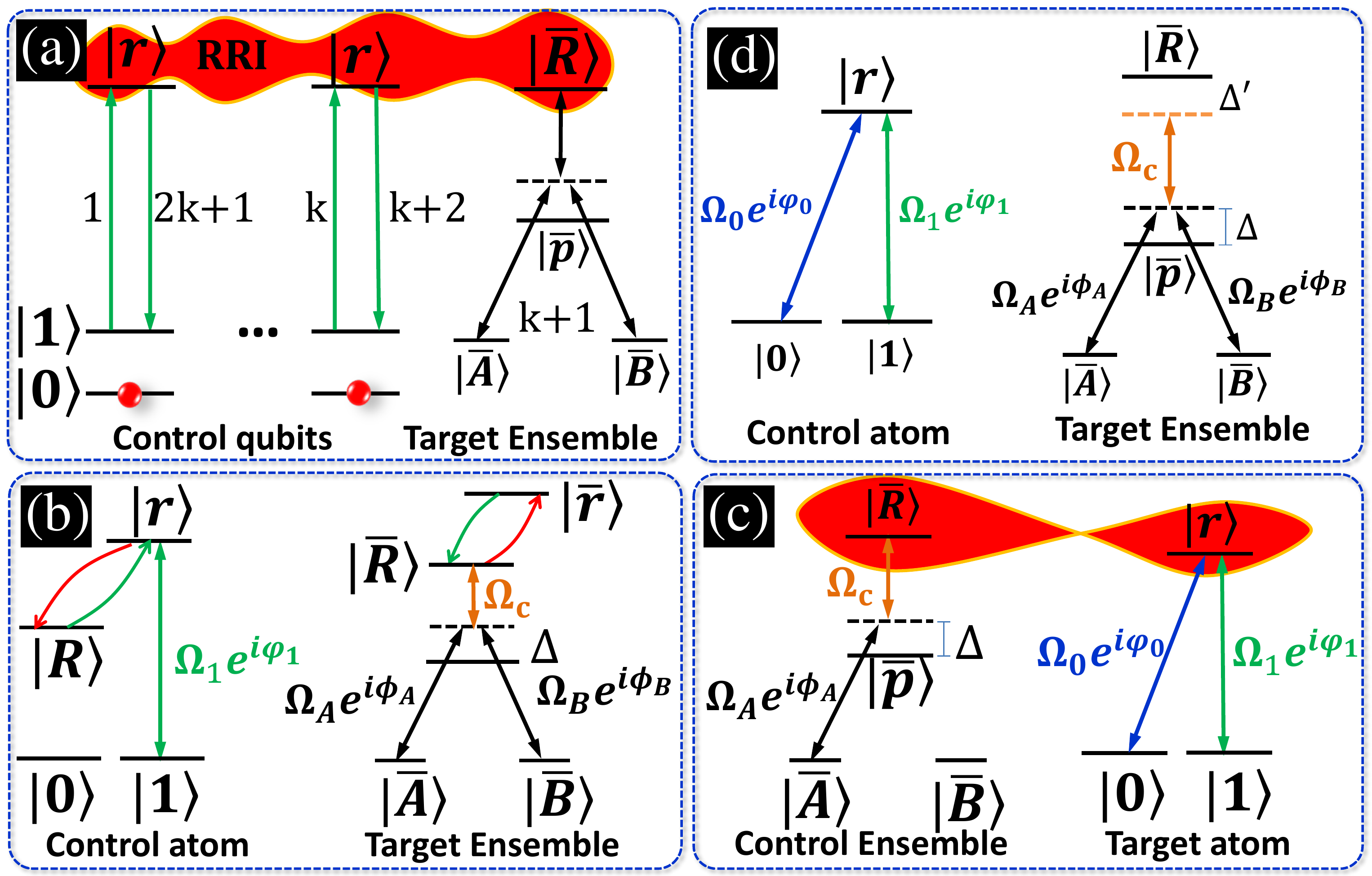}
   	\caption{[panel~(a)]. Multiple-qubit NHQC gate pulse sequence and laser driving, where the number label the order of the pulses. [panel (b)]. Energy level and laser driving of two-qubit NHQC gate based on Rydberg dark state dynamics. [panel (c)]. Energy level and laser driving of two-qubit NHQC gate, where the Rydberg ensemble acts as control qubit and single-atom acts as target qubit. [panel (d)] Energy level and laser driving to address the decoherence problem between Rydberg and ground levels.}
   	\label{fmulti}
   \end{figure}
   \subsubsection{Scalability to multiple-qubit gate}
\begin{figure}
  	\centering
  	\includegraphics[width=\linewidth]{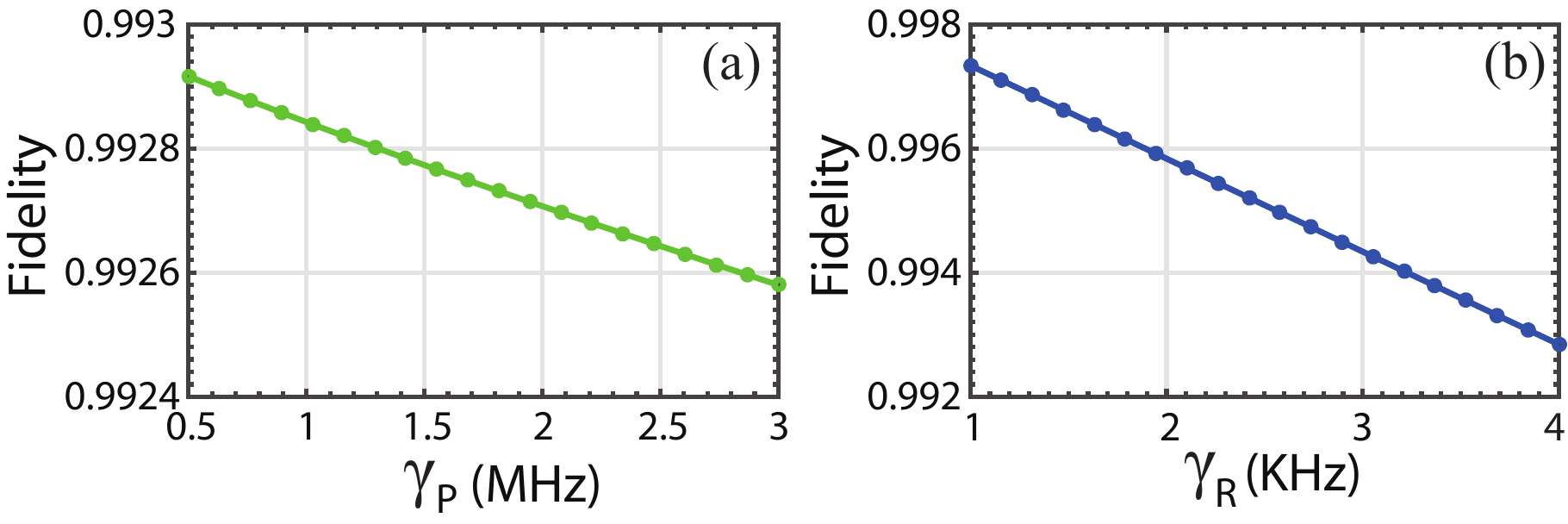}
  	\caption{Fidelity of three-qubit Toffoli gate versus decays of intermediate~[panel (a)] and Rydberg levels~[panel (b)], respectively. All of the Rabi frequencies for control atoms are the same as each other and set as $\Omega_{1} = 2\pi\times10$~MHz. The rest parameters are set as $N=4$, $\Omega_{B}=\Omega_{1}$, $\Omega_{A}=\Omega_{B}\tan(\vartheta/2)$, $\Omega_{C}=\Omega_{B}$, $\Delta=12\Omega_{B}$, $V=2\Delta$, $\phi_{A}=\phi_{B}=0$, $\mathcal{\vartheta}=-\pi/2$. The initial state is set as $(\sqrt{0.1}|00A\rangle+\sqrt{0.9}|00B\rangle+|01A\rangle+|01B\rangle+|10A\rangle+|10B\rangle+|11A\rangle+|11B\rangle)/\sqrt{7}$. For panel (a), $\gamma_{R}=\gamma_{r}=4$~KHz. For panel~(b) $\gamma_{p}=1$~MHz.}
  	\label{f009a}
  \end{figure}

   Our two-qubit geometric quantum computation scheme is able to generalize to multiple-qubit case via the conditional dynamics based on Rydberg blockade. As shown in Fig.~\ref{fmulti}(a), we consider more control atoms inspired by the basic process described in Ref.~\cite{Isenhower2011}. Suppose the RRIs between any two control atoms as well as between control atom and target ensemble are strong enough to induce the blockade effect. Thus, the geometric quantum operations on target ensemble can be performed if and only if all of the control atoms are in $|0\rangle$ state. One can achieve the evolution operator as
   	\begin{widetext}
   \begin{equation}
  \mathscr{\hat{U}}=|00\cdots0\rangle_{12\cdots k}\langle00\cdots0| \otimes\mathcal{\hat{U}}+(\sum_{a,b,\cdots m = 0,1} |ab\cdots m\rangle_{12\cdots k}\langle ab\cdots m| - |00\cdots0\rangle_{12\cdots k}\langle00\cdots0|)\otimes\mathcal{\hat{I}},
   \end{equation}
   \end{widetext}
  where $\mathcal{\hat{U}}$ denotes the Holonomic or optimized geometric operation and $\mathcal{\hat{I}}$ denote the identity matrix on the target ensemble. To verify the feasibility in a simple way, we here only consider the three-qubit Holonomic Toffoli gate. And the fidelity of the gate with one group of specified state is shown in Fig.~\ref{f009a}. 

   \subsubsection{Compatibility to other Rydberg dynamics}
\begin{figure}
 	\centering
 	\includegraphics[width=\linewidth]{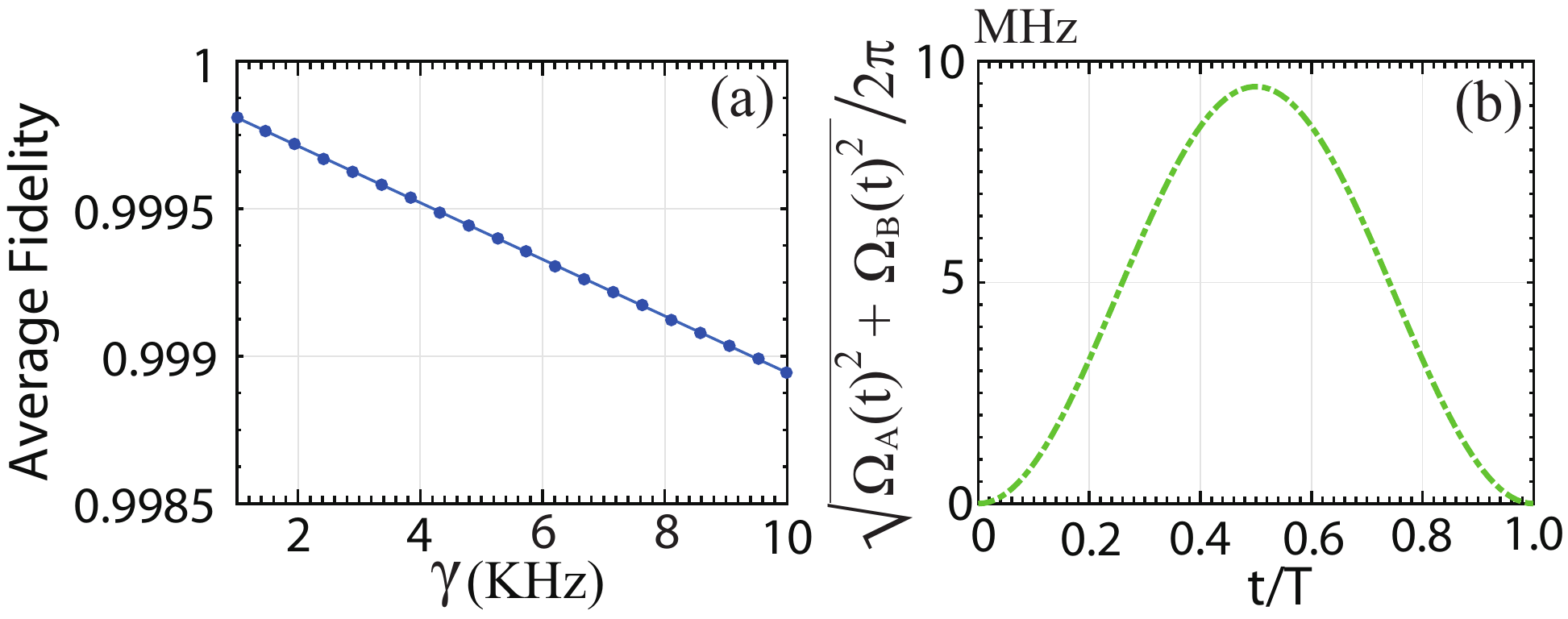}
 	\caption{Average fidelity of the scheme versus atom decay based on dark-state dynamics~[panel (a)] via the adiabatic pulse [panel (b)] in the second step. The parameters are chosen as $\Omega_{1} = 2\pi\times10$~MHz, $\varphi_{1}=0$, $N=4$, $\Omega_{A}'(t)=\Omega_{B}'(t)\tan(\vartheta/2)$, $V'=20\Omega_1$, $\phi_{A}=\phi_{B}=0$, $\vartheta=-\pi/2$, $T=0.2121\mu s$, $\sqrt{\Omega_{A}'^2+\Omega_{B}'^2} = 4\pi/T\sin(\pi t/T)^2$, and $\gamma_{R} = \gamma_{r} = \gamma$.}
 	\label{f010a}
 \end{figure}

   In above analysis, we combines the geometric phase and optimal control with Rydberg blockade to construct the quantum logic gates between single control atom and Rydberg ensemble. In this subsection, we show that the basic ideas of our Rydberg geometric quantum operations are compatible to other Rydberg dynamics. We here consider the dark-state Rydberg quantum logic gate dynamics proposed in Ref.~\cite{Petrosyan2017}. The relevant energy level and laser driving is shown in Fig.~\ref{fmulti}(b). In contrast to the conventional scheme discussed above, here we introduced one more Rydberg state $|R\rangle$ for control atom and one more state $|\bar{r}\rangle= \frac{1}{\sqrt{N}}\sum_{l=1}^{N}|a\rangle_{1}|a\rangle_{2}\cdots|r\rangle_{l}\cdots|a\rangle_{N}$ for target ensemble. The Hamiltonian for control atom is the same as Eq.~(\ref{e024n}), and for target ensemble and RRI can be written as 
   \begin{eqnarray}\label{e028}
 \hat{H}_{e} =&& \frac{1}{2}e^{i\Delta t}(\Omega_{A}e^{i\phi_{A}}|\bar{A}\rangle\langle \bar{p}|+{\Omega_{B}e^{i\phi_{B}}}|\bar{B}\rangle\langle \bar{p}|)\cr\cr&&
   	+ \frac{1}{2}e^{i\Delta t}{\Omega_{C}}|\bar{R}\rangle\langle \bar{p}|+ {\rm H.c.}
   \end{eqnarray} 
 and
    \begin{eqnarray}\label{e029}
 \hat{H}_{V} = \sum_{l=1}^{N}V_{cl}|r\rangle_{c}\langle R|\otimes|R\rangle_{l}\langle r|,
 \end{eqnarray}
respectively. In Eq.~(\ref{e029}), \emph{l} denotes the \emph{l}th atom in Rydberg ensemble and $V_{cl}$ denote the RRI strength between control atom and \emph{l}th ensemble atom. Based on the definitions of ensemble qubits, one can rewrite Eq.~(\ref{e029}) as
    \begin{eqnarray}\label{e030}
\hat{H}_{V} = V'|r\rangle_{c}\langle R|\otimes|\bar{R}\rangle\langle \bar{r}|,
\end{eqnarray}
where $V' = \sum_{l=1}^{N}V_{cl}/N$. Similar to the results in Sec.~\ref{s2.1.3}, based on the second-order perturbation theory if the condition $\Delta\gg\Omega_{A,B,C}$ and after canceling the stark shifts~\cite{note}, Eq.~(\ref{e028}) can be replaced well by the effective form 
\begin{eqnarray}\label{e031}
\hat{H}_{e} = \frac{\sqrt{\Omega_{A}'^2+\Omega_{B}'^2}e^{i\phi_{B}}}{2}|\bar{\mathcal{B}}\rangle\langle\bar{R}|+ {\rm H.c.},
\end{eqnarray}
 where $\Omega_{A}' = \Omega_{A}\Omega_{C}/(2\Delta)$ and $\Omega_{B}' = \Omega_{B}\Omega_{C}/(2\Delta)$, 
$\tan(\vartheta/2) =\Omega_{A}'/\Omega_{B}'$, $|\mathcal{\bar{B}}\rangle=\sin(\vartheta/2)e^{i(\phi_{A}-\phi_{B})}|\bar{A}\rangle+\cos(\vartheta/2)|\bar{B}\rangle$, $|\mathcal{\bar{D}}\rangle = \cos(\vartheta/2)|\bar{A}\rangle-\sin(\vartheta/2)e^{-i(\phi_{A}-\phi_{B})}|\bar{B}\rangle$. The scheme can be divided into three steps~\cite{Petrosyan2017}. The first step is to excite the control atom from $|1\rangle$ to $|r\rangle$ through $\pi$ pulse. The second step is to perform Holonomic or optimized geometric operations on the ensemble qubit. If the control atom is initially in $|0\rangle$ state, the RRI is not exist in the second step, and the geometric operation $\hat{\mathcal{U}}$ would be performed on the ensemble qubit with the laser pulse shown in Fig~\ref{f010a}(b). Otherwise, if the control atom is initially in $|1\rangle$ state, it would be excited after the first step. And the whole system Hamiltonian can be written as 
 \begin{equation}\label{e032}
\hat{H} = \frac{\sqrt{\Omega_{A}'^2+\Omega_{B}'^2}e^{i\phi_{B}}}{2}|r\bar{\mathcal{B}}\rangle\langle r\bar{R}|+ V'|r\bar{R}\rangle\langle R\bar{r}|+{\rm H.c.}.
\end{equation}
Eq.~(\ref{e032}) has one dark state 
\begin{equation}
|d\rangle = V'|r\mathcal{\bar{B}}\rangle+\frac{\sqrt{\Omega_{A}'^2+\Omega_{B}'^2}e^{i\phi_{B}}}{2}|R\bar{r}\rangle
\end{equation}
At the initial moment of the second step, the laser pulses on ensemble has not been switched on, $\Omega'_A(0) =  \Omega'_B(0) = 0$, the two-atom state $|r\mathcal{\bar{B}}\rangle$ coincides with the dark state $|d\rangle$.
During the second step, if $\sqrt{\Omega_{A}'(t)^2+\Omega_{B}'(t)^2}$ is sufficiently
smooth~[As shown in Fig~\ref{f010a}(b)], the system adiabatically follows the dark state $|d\rangle$, and the bright states that orthogonal to $|d\rangle$ would never be populated. Then, if Max$[\sqrt{\Omega_{A}'(t)^2+\Omega_{B}'(t)^2}]\ll V'$ is satisfied, the population of $|R\bar{r}\rangle$ in the dark state $|d\rangle$ could be ignored. Therefore, one can safely get that the state $|r\bar{\mathcal{B}}\rangle$ keeps invariant. In this case, the lasers has performed $\mathcal{\hat{I}}$ on the ensemble qubit. Thus, in the second step, the operations  
$\mathscr{\hat{U}}=|0\rangle_{c}\langle0|\otimes\mathcal{\hat{U}}+|r\rangle_{c}\langle r|\otimes\mathcal{\hat{I}}$
is achieved. The third step is the inverse operation of the first step. After these three steps  
\begin{equation}
\mathscr{\hat{U}}=|0\rangle_{c}\langle0|\otimes\mathcal{\hat{U}}+|1\rangle_{c}\langle1|\otimes\mathcal{\hat{I}}
\end{equation}
would be achieved.

 In Fig.~\ref{f010a}, we plot the fidelity of the controlled-NOT gate based on this dark state dynamics. One can see that the scheme has higher fidelity and is also robust on the decay of Rydberg level. We should point out that, to save the computational subspace, for the ensemble qubit we use Hamiltonian~(\ref{e031}) for simulation. And the performance may be decreased slightly if we use Eq.~(\ref{e028}). From the chosen parameters, one can easily verify $|\partial\sqrt{\Omega_{A}'(t)^2+\Omega_{B}'(t)^2}/\partial t|\ll V'^2$, which means that the adiabatic condition is satisfied well~\cite{Petrosyan2017}.
 
    \subsubsection{Exchange the roles of atom and ensemble}
\begin{figure}
      	\centering
      	\includegraphics[width=\linewidth]{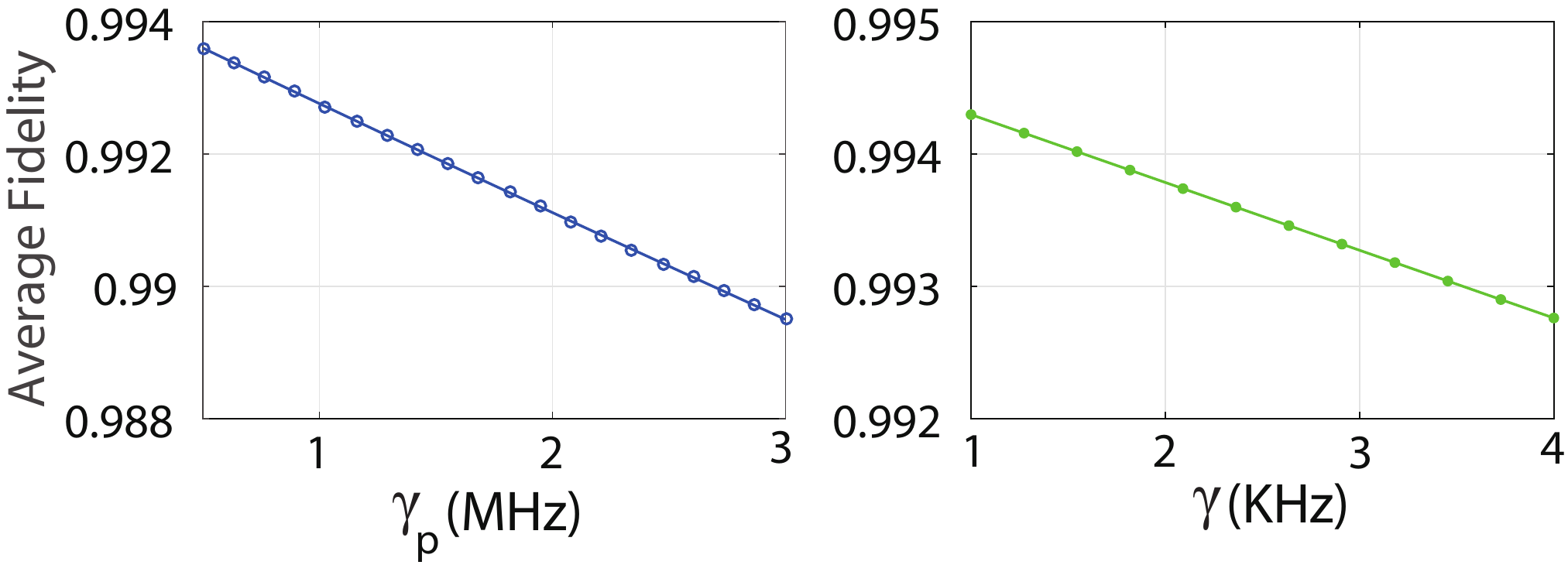}
      	\caption{Average fidelity of the scheme with Rydberg ensemble as control qubit and single-atom as target qubit, respectively. [panel (a)] Fidelity versus spontaneous emission of intermediate state. [panel (b)] Fidelity versus atomic spontaneous emission rate of Rydberg state. The parameters are chosen as $N=4$, $\Omega_{A} = 2\pi\times10$~MHz, $\Omega_{B} = 0$, $\Omega_{C} = \Omega_{A}$, $\Delta = 12\Omega_{A}$, $V=2\Delta$, $\Omega_{1}=\Omega_{A}$, $\Omega_{0} = \Omega_1\tan(\theta/2)$,  $\phi_{A}=0$, $\varphi_{0}=\varphi_{1}=0$, $\theta=-\pi/2$, and $\gamma_{R} = \gamma_{r} = \gamma$. For panel~(a), $\gamma$ is set as 4~KHz. For panel~(b), $\gamma_{p}$ is set as 1~MHz.}
      	\label{f011a}
      \end{figure}

      To show the flexibility of our scheme, we now change the roles of control and target qubits, i.e., we use Rydberg ensemble as control and single-atom as target qubits, respectively, as shown in Fig.~\ref{fmulti}(c). Similar to the schemes discussed in Sec.~\ref{s3}, three steps are also required. The first step is to excite the control ensemble qubit. The second step is to perform NHQC operations on the target single atom. If the control ensemble is excited, the target single-atom operations would be inhibited. The third step is to deexcite the control ensemble. The fidelity is shown in Fig.~\ref{f011a}, which shows the scheme is also feasible after changing the roles of single Rydberg atom and ensemble.
      
     \subsection{Deal with some imperfections  of Rydberg ensemble}

      The dominant factor that influences the applications of Rydberg ensemble is the decoherence problem between the ground and Rydberg levels~\cite{mmtm2015,Zeiher2015,Ebert2014,tmtt2015,Saffman_2016}. This is because the Rydberg ensemble may be sensitive to field gradients as well as the presence of atomic collisions and possibly molecular resonances~\cite{Derevianko2015}. These effects may be mitigated by the method as described in Ref.~\cite{Saffman2008}. In this subsection, we would give another method to address this issue by adiabatically canceling the Rydberg state, and the relevant laser driving is shown in Fig.~\ref{fmulti}(d). The Hamiltonian for single control atom is the same as Eq.~(\ref{e04}) if we set $\Omega_{0} = 0$ and $\varphi_{1}=0$. The Hamiltonian for ensemble qubit is redesigned as 
         \begin{eqnarray}\label{e035}
       \hat{H}_{e} =&& \frac{1}{2}e^{i\Delta t}(\Omega_{A}e^{i\phi_{A}}|\bar{A}\rangle\langle \bar{p}|+{\Omega_{B}e^{i\phi_{B}}}|\bar{B}\rangle\langle \bar{p}|)\cr\cr&&
      + \frac{1}{2}e^{i(\Delta+\Delta') t}{\Omega_{C}}|\bar{R}\rangle\langle \bar{p}|+ {\rm H.c.}.
      \end{eqnarray}
      Similar to the process from Eq.~(\ref{e028}) to Eq.~(\ref{e031}), after adiabatically canceling the $|\bar{p}\rangle$ state and  some relevant stark shifts, one can get the effective form as
           \begin{eqnarray}\label{e036}
        \hat{H}_{e} = \frac{\sqrt{\Omega_{A}'^2+\Omega_{B}'^2}e^{i\phi_{B}}}{2}|\bar{\mathcal{B}}\rangle\langle\bar{R}|e^{-i\Delta' t}+ {\rm H.c.},
        \end{eqnarray}
        If $\Delta'=0$, the Hamiltonian is back up to the form in Eq.~(\ref{e07n}). And the corresponding two-qubit gate is the same as that discussed in Sec.~\ref{s3.3}. Here, to address the issue of the decoherence between Rydberg and ground levels, we consider the dispersive regime with the condition $\Delta'\gg\{\Omega_{A}',\Omega_{B}'\}$ and consider both of the decay and dephasing rates. 
        Then, Eq.~(\ref{e036}) is simplified to the effective form
         \begin{eqnarray}\label{e037}
        \hat{H}_{e}^{\rm eff} =- \frac{\Omega_{A}'^2+\Omega_{B}'^2}{4\Delta'}|\bar{\mathcal{B}}\rangle\langle\bar{\mathcal{B}}|,
        \end{eqnarray}
        if the state is initially in ground state subspace. From the perspective of NHQC, one can get that Eq.~(\ref{e037}), i.e., the effective form of Eq.~(\ref{e036}), satisfies the cyclic condition automatically. From the perspective of robustness, the Rydberg levels have been canceled in Eq.~(\ref{e037}), which means the decoherence may be decreased~(The numerical demonstration would be given later). To construct the NHQC gate based on the dispersive regime of the ensemble qubit, three steps that similar to the processes in Sec.~\ref{s3.3} are required. The difference is that the evolution time \emph{T} should be decided by $\int_{0}^{T}(\Omega_{A}'^2+\Omega_{B}'^2)/(2\Delta')dt=2\pi$ in the second step. 
    
\begin{figure}
  	\centering
  	\includegraphics[width=\linewidth]{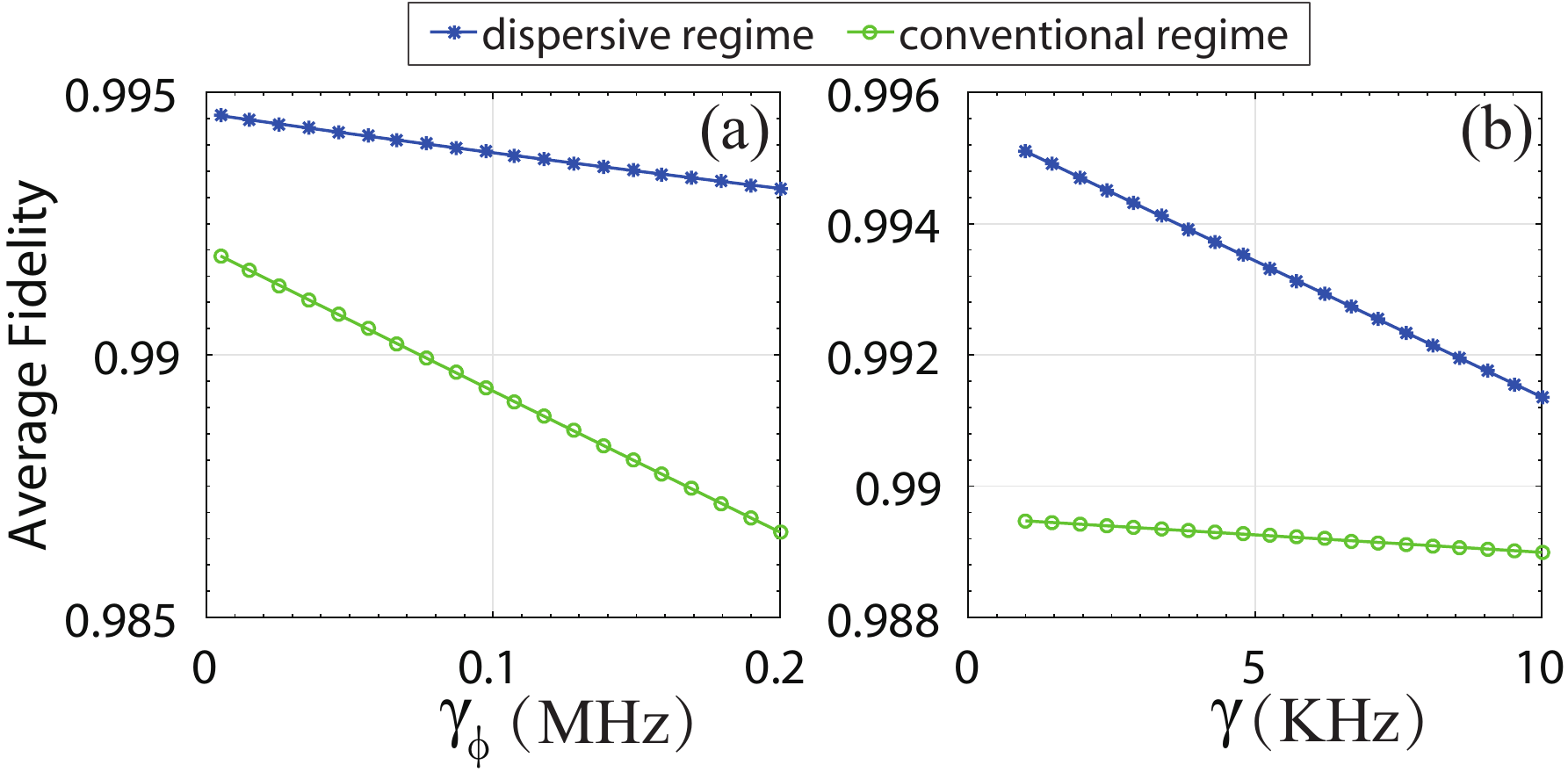}
  	\caption{Average fidelity versus decoherence between the Rydberg and ground levels. [Panel (a)] Versus dephasing rate. [Panel (b)] Versus spontaneous emission rate. Green circle lines denote the conventional regime that we discussed previously. Blue star lines denote the dispersive regime that we introduced to address the issues of the decoherence problem between the Rydberg and ground levels. The parameters are set as $\Omega_{1} = 2\pi\times10$~MHz, $\varphi_{1}=0$, $N=4$, $\Omega_{B}'=\Omega_{1}$, $\Omega_{A}'=\Omega_{B}'\tan(\vartheta/2)$, $\Delta'=10\Omega_{B}'$, $V=20\Delta'$, $\phi_{A}=\phi_{B}=0$, $\mathcal{\vartheta}=-\pi/2$. For panel (a), $\gamma_{R}=\gamma_{r}=4$~KHz. For panel~(b) $\gamma_{R}=\gamma_{r}=\gamma$, $\gamma_{\phi}= 100$~KHz.}
  	\label{f012a}
  \end{figure}

      In Fig.~\ref{f012a}, we plot the average fidelity of the NHQC gate based on the conventional and dispersive regimes, respectively. Panel~(a) and (b) show the average fidelity versus the dephasing rate and decay, respectively. In Fig.~\ref{f012a}(a), the dephasing operator for Rydberg ensemble is defined as $\mathcal{\hat{L}_\phi}=\sqrt{\gamma_{\phi}}(\mathcal{\hat{I}}-2|\bar{R}\rangle\langle\bar{R}|)$~\cite{Rao2014}. One can see that the average fidelity of dispersive regime is higher than that of the conventional regime with the consideration of dissipation. In other words, this proposed dispersive regime reduced the influence of the decoherence between Rydberg and ground levels. 
  
  Besides, for the conventional encoding method, although the excitation Rabi frequency could be enhanced by a factor $\sqrt{N}$, the construction of high fidelity gate operations may be problematic when \emph{N} is not accurately known~\cite{Saffman_2016}. In this manuscript, we encoded the ensemble atom inspired by the encoding method in Refs.~\cite{imed2013, mmtm2015}. One can calculate that the Rabi frequency is independent of ensemble atom number $N$ according to the expressions in Eq.~(\ref{e021j}).  

	\subsection{Experimental considerations}
\begin{figure}
	\centering
	\includegraphics[width=\linewidth]{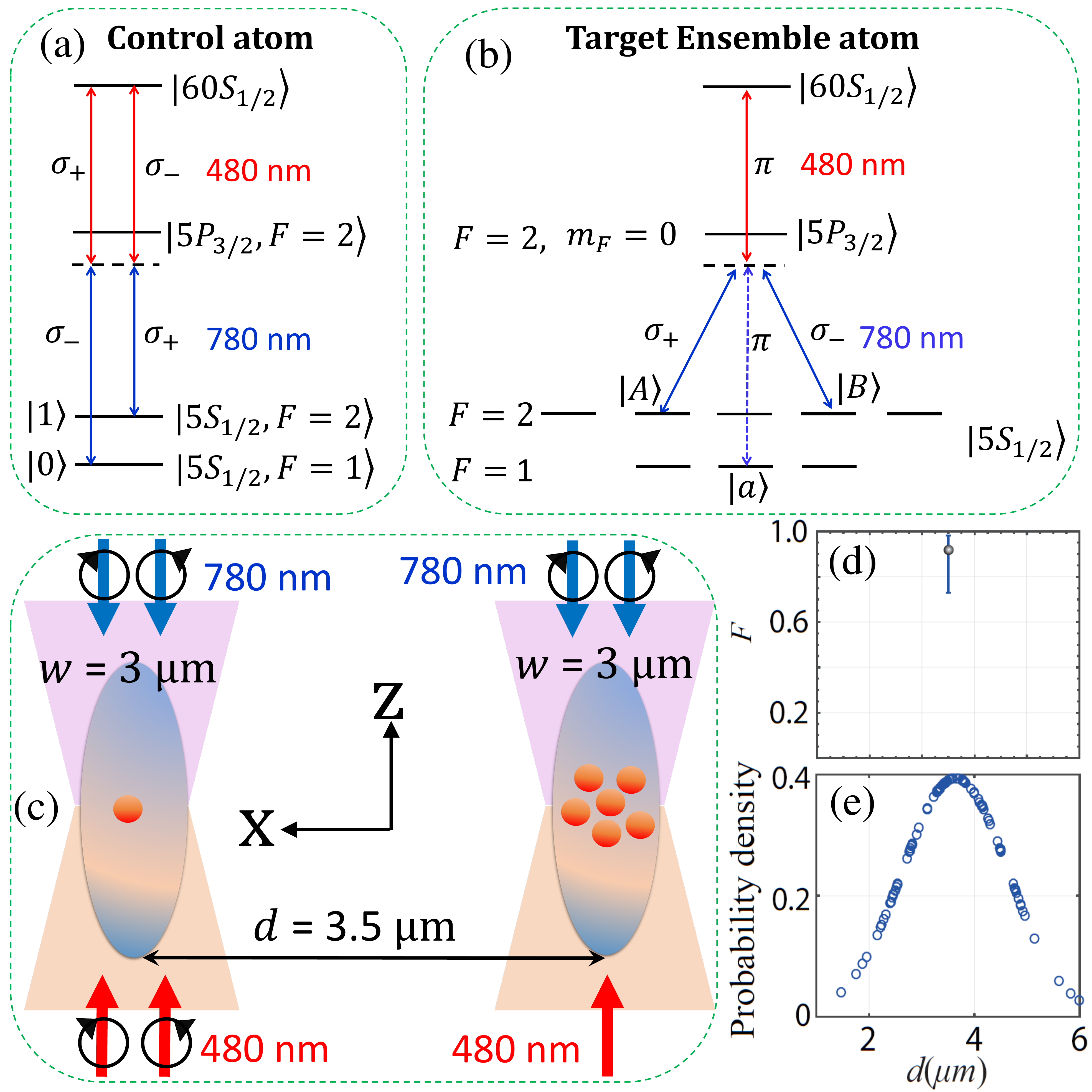}
	\caption{Energy levels of the control atom [panel~(a)] and target ensemble atom~[panel (b)], respectively. The Rydberg excitation are enabled by a two-color laser system at 780~nm and 480~nm. For the control atom, the 780~nm $\sigma_{+}$ pulse couples $|1\rangle$ with intermediate state, which further couples to the Rydberg state with 480~nm $\sigma_{-}$ pulse. And the 780~nm $\sigma_{-}$ pulse couples $|0\rangle$ with the intermediate state, which further couples to the Rydberg state with 480~nm $\sigma_{+}$ pulse. For the ensemble atom, three 780~nm pulses with $\sigma_{+}$, $\sigma_{-}$, and $\pi$ polarization couple $|A\rangle$, $|B\rangle$ and $|a\rangle$ to intermediate state, respectively. The $|p\rangle$ state couples to Rydberg state via 480~nm pulse with $\pi$ polarization. The dashed line means that the pulse only turned on in the initial ensemble qubit preparation process. (c) Experimental geometry with ideal inter-atomic distance $d=3.5~\mu$m laser optical waist $w=3~\mu$m. (d) Errorbar of the average fidelity with the consideration of one group of random number which satisfies Gaussian distribution with average value $3.5~\mu m$ and standard deviation $0.9~\mu m$~[as shown in panel (e)]. In panel~(d), the rest parameters are given in Table~\ref{t1}.}
	\label{f008}
   \end{figure}

	To implement our two-qubit scheme experimentally, we consider Rb atoms and the relevant energy levels are shown in Fig.~\ref{f008}. To be concrete, one can choose $|0\rangle\equiv|5S_{1/2},F=1,m_{F}=0\rangle$, $|1\rangle\equiv|5S_{1/2},F=2,m_{F}=0\rangle$, $|r\rangle=|60S_{1/2},F=1,m_{F}=0\rangle$ for the control atom. The intermediate energy for two-photon process can be chosen as $|5P_{3/2},F=2\rangle$ and $m_{F}=-1(+1)$ for $|0(1)\rangle\rightarrow|r\rangle$ process. For target ensemble atoms, energy level are chosen as $|a\rangle\equiv|5S_{1/2},F=1,m_{F}=0\rangle$, $|A\rangle\equiv|5S_{1/2},F=2,m_{F}=-1\rangle$, $|B\rangle\equiv|5S_{1/2},F=2,m_{F}=+1\rangle$, $|p\rangle\equiv|5P_{3/2},F=2,m_{F}=0\rangle$, $|R\rangle=|60S_{1/2},F=1,m_{F}=0\rangle$. The Rydberg excitations are enabled by a two-color laser system at 780~nm and 480~nm. For the 780~nm laser, it can be modulated with an acousto-optic modulator~(AOM) driven by an arbitrary waveform generator~(AWG) to achieve the effective Rabi frequency shape~\cite{Omran570,Higgins2017}. Meanwhile, the desired laser phase can also be modulated through changing  phases of the radio-frequency drive of the AOM~\cite{Omran570}. More importantly, the optimized pulse obtained by different optimal method has been employed to prepare multiple Rydberg atom entangled Greenberger-Horne-Zeilinger state in Ref.~\cite{Omran570}, in which the similar experiment pulse configuration is useful to the experimental implementation of our scheme.

		\renewcommand\tabcolsep{10.5pt}
		\begin{table*}[htp!]
			\centering \caption{Average fidelity and whole evolution time of two-qubit controlled-NOT gate versus the optimization parameter $n$ (ensemble atom number $N = 4$). Max[$\Omega_{1}(t)/2\pi] =\max[\Omega_{\rm eff}(t)/2\pi]$= 6~MHz, $\Omega_{C}/2\pi=140$~MHz, $\Delta/2\pi=2$~GHz, $\gamma_{r}=\gamma_{R} = 4.4$~kHz and $\gamma_{p} = 38$~MHz.}
			\begin{tabular}{lcccccccccc}
				\hline\hline
				$n$&0.1&0.2&0.3&0.4&0.5&0.6&0.7&0.8&0.9&1.0\\ \hline
				$T$~(ns)&359.01&426.88&520.68&628.93&745.36&866.67&991.07&1117.5&1245.4&1374.4\\
				$F$&0.9641 &0.9662 & 0.9685&0.9730 &0.9782&0.9823&0.9846&0.9852& 0.9841& 0.9820\\
				\hline
			\end{tabular}\label{t1}
		\end{table*}

	The inter-atomic distance among target ensemble atoms may be less than the characteristic length $R_{c}$~\cite{mtk2010}, and thus the dipole-dipole-interaction-induced blockade would play the main role in the ensemble. While for the control-target interaction, any one of the dipole-dipole-interaction-induced blockade and vdW-interaction induced blockade may be feasible for our scheme. $C_{6}$ between $|r\rangle$ and $|R\rangle$ are about $139$~GHz$\cdot\mu m^6$~\cite{Li2014,singer2005long} for our chosen level. If the average distance between control and target ensemble atom are set as 3.5~$\mu m$, the value of RRI is about $V/2\pi=75.6$~MHz. And for the chosen level, $\gamma_{r}=\gamma_{R}\simeq4.4$~kHz~\cite{Rydbergdecay}, $\gamma_{p}\simeq38$~MHz~\cite{steck2001rubidium}. If the way to design laser pulse is the same as that in Sec.~\ref{s3.3} but with max[$\Omega_{1}(t)/2\pi] =\max[\Omega_{\rm eff}(t)/2\pi]$= 6~MHz, $\Omega_{C}/2\pi=140$~MHz and $\Delta/2\pi=2$~GHz, the fidelity of two-qubit controlled-NOT gate with ensemble atom number $N=4$ would still be about 0.985 even when the systematic error reaches 10\%. The optimal parameter $n=0.7$ is set and the whole time of the gate can reach submicroscopic magnitude~(991.07~ns). To further shorten the evolution time, one can decrease \emph{n} with the price of reducing the optimization effect (As shown in Table.~\ref{t1}).

  Typical beams powers of  $2.3~\mu$W at 780 nm and of $12$~mw at 480 nm beam are employed to achieve the two-photon process $5s_{1/2}\xrightarrow[2.3~\mu W]{\rm 780~nm}5p_{3/2}\xrightarrow[\rm12~mW]{\rm 480~nm}97d_{5/2}$ with the intermediate state detuning about 1.1 GHz~\cite{lex2010}. The resulted $\pi$ Rydberg pulse times of 750 ns with laser waist $w=10~\mu$m~\cite{lex2010,Gaetan2009}. And one can inversely calculate the effective Rabi frequency as $\max[\Omega(t)/2\pi]=0.667$~MHz. The Rabi frequency is relevant to the electric field $\bm{E}$ and electric dipole moment $\bm{d}$ as~\cite{scully1999quantum}
  \begin{equation}
 \Omega=\frac{\bm{d}\cdot \bm{E}}{\hbar}.
  \end{equation}
 For Gaussian beams, optical intensity $I\propto |\bm{E}|^2$, optical power $P\propto IS$, area $S\propto w^2$ where $w$ denotes the laser beam waist. one can change these two parameters to enhance the effective Rabi frequency. The dipole matrix of our scheme for the transition $5s_{1/2}\rightarrow5p_{3/2}$ is the same as that of Ref.~\cite{lex2010}. For the other transition in the two-photon process, $5p_{3/2}\rightarrow60s_{1/2}$ is employed in our scheme while in Ref.~\cite{lex2010} $5p_{3/2}\rightarrow97d_{5/2}$ is employed, which means the dipole moment $\bm{d}$ is different. $\bm{d}$ can be reduced as radial matrix element and Angular matrix element and thus be calculated~\cite{steck2001rubidium, nguyen:tel-01560901,*hovanessian1976computational,*brink1968angular}.
 Thus, one can roughly evaluate that, with the same optical parameter as in Ref.~\cite{lex2010}, the effective Rabi frequency would be $\Omega/2\pi\simeq0.705$~MHz, which is far less than the max value 6~MHz of our scheme. Based on above analysis, if we reconsider the optical waist to be $3~\mu$m, and set the optical power for two-photon process as $0.39~\mu$W and $60$~mW, respectively, the desired Rabi frequency can be achieved. It should be noted that for atomic ensemble, the optical waist should be enlarged to ensure all of the atoms being illuminated, which means the optical power should be enhanced to guarantee the set Rabi frequency. In fact, in Ref.~\cite{Omran570}, the max value of 5~MHz of time-dependent Rabi frequency has been experimentally implemented.

 We now consider the influence of the position probability distributions. For simplicity, we suppose the inter-atomic distance probability distributions are approximately Gaussian with standard deviation 0.9~$\mu m$, which is about 25\% of the set inter-atomic distance $d=3.5~\mu m$. Suppose the $C_{6}$ parameter keeps invariant and the RRI only influenced by the inter-atomic distance. We plot the average fidelity with errorbar in Fig.~\ref{f008}(d). When the random inter-atomic probability density [Fig.~\ref{f008}(e)] is considered, the average fidelity is still large. Parameters that are need to realize the atomic distance can be achieved within current experiments~\cite{mmtm2015,lex2010}.

\section{Conclusions}\label{s6}
In conclusion, we have proposed schemes to implement universal quantum logic gates with mesoscopic ensembles of Rydberg atoms being the target ensemble qubit. Two related but different schemes, NHQC  and NHQC+, are examined in detail. We have shown that gate fidelities are high in our scheme. In particular, we have applied the dynamical-invariant-based optimized method to re-design the laser pulses to enhance the performance of the scheme. Our numerical results show that the optimized schemes are robust with regard to systematic errors (i.e. laser parameters) even when the laser Rabi frequency has a fluctuation as high as 10\%. Moreover we have shown through numerical simulations that the optimized method can reduce the error caused by higher-order perturbation terms. Based on practical parameters, we have demonstrated that the two-qubit gates can be implemented in submicroseconds while still achieve relatively high gate fidelities. This gate time is comparable to state-of-the-art results~\cite{Rydbergion2020}. Our proposal shows the potential to achieve scalable quantum computation with strong and controllable Rydberg interactions~\cite{Rydbergion2020}, and hence will attract future studies of underlying questions. Our study opens a new route to realize fast and robust holonomic quantum computation with mesoscopic Rydberg atom ensembles. It will contribute to the ongoing effort in developing quantum simulation and computation with Rydberg atoms.

\emph{Acknowledgement}---
We would like to thank Dr. B. J. Liu for useful discussions. This work was supported by National Natural Science Foundation of China (NSFC) under Grant Nos. 11804308 and 11804375. And China Postdoctoral Science Foundation (CPSF) under Grant No. 2018T110735. W. L. acknowledges support from the EPSRC through grant No. EP/R04340X/1 via the QuantERA project “ERyQSenS”, the Royal Society grant No. IEC$\backslash$NSFC$\backslash$181078, and  the UKIERI-UGC Thematic Partnership No. IND/CONT/G/16-17/73.

	\appendix
	\renewcommand{\thefigure}{A\arabic{figure}}
	\section{Preparation of ensemble qubit states} \label{s7}

	As shown in Fig.~$\ref{f001}$, we consider the mesoscopic atom ensemble consists of \emph{N} identical five-level atoms, each of which has three ground states $|A\rangle$, $|a\rangle$ and $|B\rangle$, an intermediate state $|p\rangle$ and a Rydberg state $|R\rangle$. In this case, we use the ground state $|a\rangle$ to generate the collective states that we need.
	Suppose all of the ensemble atoms are prepared in state $|a\rangle$, i.e., the initial ground collective state $|\overline{a}\rangle=|a_{1}\cdots a_{N}\rangle$, where $|a_{l}\rangle$ represents the \emph{l}-th Rydberg atom is in the ground state $|a\rangle$. Then, we employ laser to couple auxiliary state $|a\rangle$ to Rydberg state $|R\rangle$. $|\overline{R}\rangle$ would be generated because of Rydberg blockade. Then we drive the Rydberg atom from the state $|R\rangle$ to $|A\rangle$ or $|B\rangle$ or $|p\rangle$, where we prepare a collective Rydberg state, an intermediate state and two collective ground states:
	\begin{eqnarray}\label{e021j}
	|\overline{R}\rangle &=& \frac{1}{\sqrt{N}} \sum_{l=1}^{N}|a\rangle_{1}|a\rangle_{2}\cdots|R\rangle_{l}\cdots|a\rangle_{N}\cr\cr
	|\overline{p}\rangle &=& \frac{1}{\sqrt{N}}
	\sum_{l=1}^{N}|a\rangle_{1}|a\rangle_{2}\cdots|p\rangle_{l}\cdots|a\rangle_{N}\cr\cr
	|\overline{A}\rangle &=& \frac{1}{\sqrt{N}} \sum_{l=1}^{N}|a\rangle_{1}|a\rangle_{2}\cdots|A\rangle_{l}\cdots|a\rangle_{N}\cr\cr
	|\overline{B}\rangle &=& \frac{1}{\sqrt{N}}\sum_{l=1}^{N}|a\rangle_{1}|a\rangle_{2}\cdots|B\rangle_{l}\cdots|a\rangle_{N}
	\end{eqnarray}
	where footnote $l$ denotes the \emph{l}-th atom.
	
	\section{Derivations of $\dot{\Theta}$, ~$\dot{\alpha}$, and $\dot{\gamma}$  }\label{s8}
	\subsection{method one}

Due to the Hermitian operator $\hat{I(t)}$ satisfies the  $\frac{\partial}{\partial t}\hat{I(t)}+i[\hat{\mathcal {H}}_{\rm eff},\hat{I}]=0$.
So we can get Eq.~(\ref{e40j}) by takng the Eqs.~(\ref{e20j}) and (\ref{e21j}) into this formula:
\begin{widetext}
\begin{eqnarray}\label{e40j}
 \left(
  \begin{array}{cccc}
   -\sin\Theta(\dot{\Theta} +\Omega_{R}\sin\alpha-\Omega_{I}\cos\alpha)  & i\cos\Theta(-\Omega_{R}+i\Omega_{I})+(-ie^{-i\alpha}\dot{\alpha}\sin\Theta+e^{-i\alpha}\dot{\Theta}\cos\Theta) \\
     i\cos\Theta(\Omega_{R}+i\Omega_{I})+   (ie^{i\alpha}\dot{\alpha}\sin\Theta+e^{i\alpha}\dot{\Theta}\cos\Theta) & \sin\Theta(\dot{\Theta}+\Omega_{R}\sin\alpha-\Omega_{I}\cos\alpha)\\
  \end{array}
\right)=0\cr\cr
\end{eqnarray}\end{widetext}

Thus :
\begin{eqnarray}\label{e27j}
 &&-\sin\Theta(\dot{\Theta}+\Omega_{R}\sin\alpha-\Omega_{I}\cos\alpha)=0\cr\cr
 &&(-ie^{-i\alpha}\dot{\alpha}\sin\Theta+e^{-i\alpha}\dot{\Theta}\cos\Theta)+i\cos\Theta(-\Omega_{R}+i\Omega_{I})=0\cr\cr
 &&(ie^{i\alpha}\dot{\alpha}\sin\Theta+e^{i\alpha}\dot{\Theta}\cos\Theta)+i\cos\Theta(\Omega_{R}+i\Omega_{I})=0
\end{eqnarray}
so we can get the value of $\dot{\Theta}$ and $\dot{\alpha}$ by solving the equation Eq.~(\ref{e27j}).

Also,
\begin{eqnarray}\label{e28j}
\dot{\gamma}&&=-2\dot{f_{+}}=-2\langle\phi_{+}|i\frac{\partial}{\partial t}-\hat{\mathcal{H}}_{\rm eff}|\phi_{+}(t)\rangle\cr\cr&&
=-2\left[\frac{\dot{\alpha}\cos\Theta}{2}-\frac{\sin\Theta(\Omega_{R}\cos\alpha+\Omega_{I}\sin\alpha)}{2}\right]\cr\cr&&
=\frac{(\Omega_{R}\cos\alpha+\Omega_{I}\sin\alpha)}{\sin\Theta}
\end{eqnarray}
There we obtain the value of $\dot{\Theta}$, $\dot{\alpha}$ and $\dot{\gamma}$.

	\subsection{method two}

	Due to the $|\psi(t)\rangle$ satisfies the Schr{\"o}dinger equation: $i\frac{\partial}{\partial t}|\psi(t)\rangle=\hat{\mathcal{H}}_{\rm eff}|\psi(t)\rangle$. So putting the Eq.~(\ref{e23}) in text into the Schr{\"o}dinger equation, one can get:
    \begin{eqnarray}\label{e29j}
    &&\left(
    \begin{array}{cccc}
    -i\sin(\Theta/2)\dot{\Theta}+\cos(\Theta/2)(\dot{\alpha}+\dot{\gamma}) \\
    i\cos(\Theta/2)\dot{\Theta}-\sin(\Theta/2)(\dot{\alpha}-\dot{\gamma}) \\
    \end{array}
    \right)\cr\cr&&=
    \left(
    \begin{array}{cccc}
    (\Omega_{R}-i\Omega_{I})\sin(\Theta/2)e^{i\alpha} \\
    (\Omega_{R}+i\Omega_{I})\cos(\Theta/2)e^{-i\alpha} \\
    \end{array}
    \right)
    \end{eqnarray}

Simplify the Eq.~(\ref{e29j}), one can get:
\begin{eqnarray}\label{e30j}
-i\sin(\Theta/2)\dot{\Theta}&=&i\sin(\Theta/2)(\Omega_{R}\sin\alpha-\Omega_{I}\cos\alpha)\cr\cr
\cos(\Theta/2)(\dot{\alpha}+\dot{\gamma})&=&\sin(\Theta/2)(\Omega_{R}\cos\alpha+\Omega_{I}\sin\alpha)\cr\cr
\sin(\Theta/2)(\dot{\gamma}-\dot{\alpha})&=&\cos(\Theta/2)(\Omega_{R}\cos\alpha+\Omega_{I}\sin\alpha)
\end{eqnarray}
We can easily get the value of $\dot{\Theta}$, $\dot{\alpha}$ and $\dot{\gamma}$ by solving the equation Eq.~(\ref{e30j}).

	\section{Pulse Expressions}\label{appc1}
	\subsection{Invariant case}
	We can obtain $\dot{\Theta}(t)=\Omega_{I}\cos\alpha-\Omega_{R}\sin\alpha$, $\dot{\alpha}=-\cot\Theta(\cos\alpha\Omega_{R}+\sin\alpha\Omega_{I})$ and
    $\dot{\gamma}=(\cos\alpha\Omega_{R}+\sin\alpha\Omega_{I})/\sin\Theta$ using method one or method two.
    Then for the given value of the $\dot{\Theta}$, $\dot{\alpha}$ and $\dot{\gamma}$, we can get the pulse expressions:
	\begin{eqnarray}\label{e25j}
	&&\Omega_{R}=\cos\alpha\sin\Theta\dot{\gamma}-\sin\alpha\dot{\Theta}\cr\cr
	&&\Omega_{I}=\sin\alpha\sin\Theta\dot{\gamma}+\cos\alpha\dot{\Theta}\cr\cr
	&&\alpha = -\int dt\cos[\Theta(t)]\dot{\gamma}(t)
	\end{eqnarray}

	\subsection{Invariant-based optimal control}
	For the given expression of $\gamma(t)=n[2\Theta-\sin(2\Theta)]$, we can get $\dot{\gamma}(t)=4n\dot{\Theta}\sin^{2}\Theta$. Then put in into the Eq.~(\ref{e25j}), we can get
\begin{eqnarray}\label{e26j}
&&\Omega_{R}=(4n\cos\alpha\sin^{3}\Theta-\sin\alpha)\dot{\Theta},\cr\cr
&&\Omega_{I}=(4n\sin\alpha\sin^{3}\Theta+\cos\alpha)\dot{\Theta}\cr\cr
&&\alpha = -4n\int dt\dot{\Theta}(t)\cos[\Theta(t)]\sin[\Theta(t)]^2 .
\end{eqnarray}
	\section{Proof of satisfying von Neumann equation}\label{s9}

	Due to the $|\psi(t)\rangle$ satisfies the Schr{\"o}dinger equation: $i\frac{\partial}{\partial t}|\psi(t)\rangle=\hat{\mathcal{H}}_{\rm eff}|\psi(t)\rangle$. Also, $-i\frac{\partial}{\partial t}\langle\psi(t)|=\langle\psi(t)|\hat{\mathcal{H}}_{\rm eff}$.
 \begin{eqnarray}\label{e32j}
i\frac{\partial}{\partial t}(|\psi(t)\rangle\langle\psi(t)|)
&&=i\frac{d}{dt}|\psi(t)\rangle\langle\psi(t)|+|\psi(t)\rangle i\frac{d}{dt}\langle\psi(t)|\cr\cr
&&=(\hat{\mathcal{H}}_{\rm eff}|\psi(t)\rangle)\langle\psi(t)|-|\psi(t)\rangle(\langle\psi(t)|\hat{\mathcal{H}}_{\rm eff})\cr\cr
&&=[\hat{\mathcal{H}}_{\rm eff},|\psi(t)\rangle\langle\psi(t)|]
\end{eqnarray}
So the $|\psi(t)\rangle\langle\psi(t)|$ satisfies the von Neumann equation. What's more, according to the Eq.~(\ref{e23}) $|\phi_{+}(t)\rangle\langle\phi_{+}(t)|=|\psi(t)\rangle\langle\psi(t)|$, so the $|\phi_{+}(t)\rangle$ ($|\nu_{+}(t)\rangle$) satisfies the von Neumann equation too.

	\section{Proof of the integral of dynamic term is zero}\label{s10}
		\subsection{Analytical results}
We have known that $\Omega_{R}=-\cos\alpha\sin\Theta\dot{\gamma}+\sin\alpha\dot{\Theta},\Omega_{I}=\sin\alpha\sin\Theta\dot{\gamma}+\cos\alpha\dot{\Theta}$. Suppose $\Omega_{R}$ suddenly have a minus sign at the half moment of the evolution, after analysis we get the value of the $\Theta, \dot{\Theta}$ keep invariant and $\gamma, \dot{\gamma}, \alpha$ have a minus sign. Then:
\begin{eqnarray}\label{e33j}
&&\int_{0}^{\tau}\langle\nu_{1}(t)|\hat{\mathcal{H}}_{\rm eff}|\nu_{1}(t)\rangle dt \cr\cr&=&\frac{1}{2}\int_{0}^{\tau}\sin\Theta(\Omega_{I}\sin\alpha+\Omega_{R}\cos\alpha)dt\cr\cr
&=&\frac{1}{2}\int_{0}^{\tau/2}\sin^{2}\Theta\dot{\gamma}dt+\frac{1}{2}\int_{\tau/2}^{\tau}-\sin^{2}\Theta\dot{\gamma}dt
\end{eqnarray}
Then, if the area enclosed by the curve of function $\sin^{2}\Theta\dot{\gamma}$ and the t-axis in the interval (0, $\tau/2$) is equal to that in the interval ($\tau/2,~\tau$),
Eq.~(\ref{e33j}) equals zero. One simple case is that the function $\sin^{2}\Theta\dot{\gamma}$ is symmetry with respect to $t = \tau/2$ axis.

For the pulses in Sec.~\ref{s4.1}, one can easily get $\int_{0}^{\tau/2}\sin^{2}\Theta\dot{\gamma}dt=\int_{\tau/2}^{\tau}\sin^{2}\Theta\dot{\gamma}dt=\pi/2$, which means Eq.~(\ref{e33j}) equals zero. And for the pulse in Sec.~\ref{slogicalqubit}, one can also demonstrate Eq.~(\ref{e33j}) equals zero since $\int_{0}^{\tau/2}\sin^{2}\Theta\dot{\gamma}dt=\int_{\tau/2}^{\tau}\sin^{2}\Theta\dot{\gamma}dt=3n\pi/4$. For two-qubit gate, similar proof process can also be given.

\setcounter{figure}{0}
\begin{figure}[htp!]
	\centering
	\includegraphics[width=\linewidth]{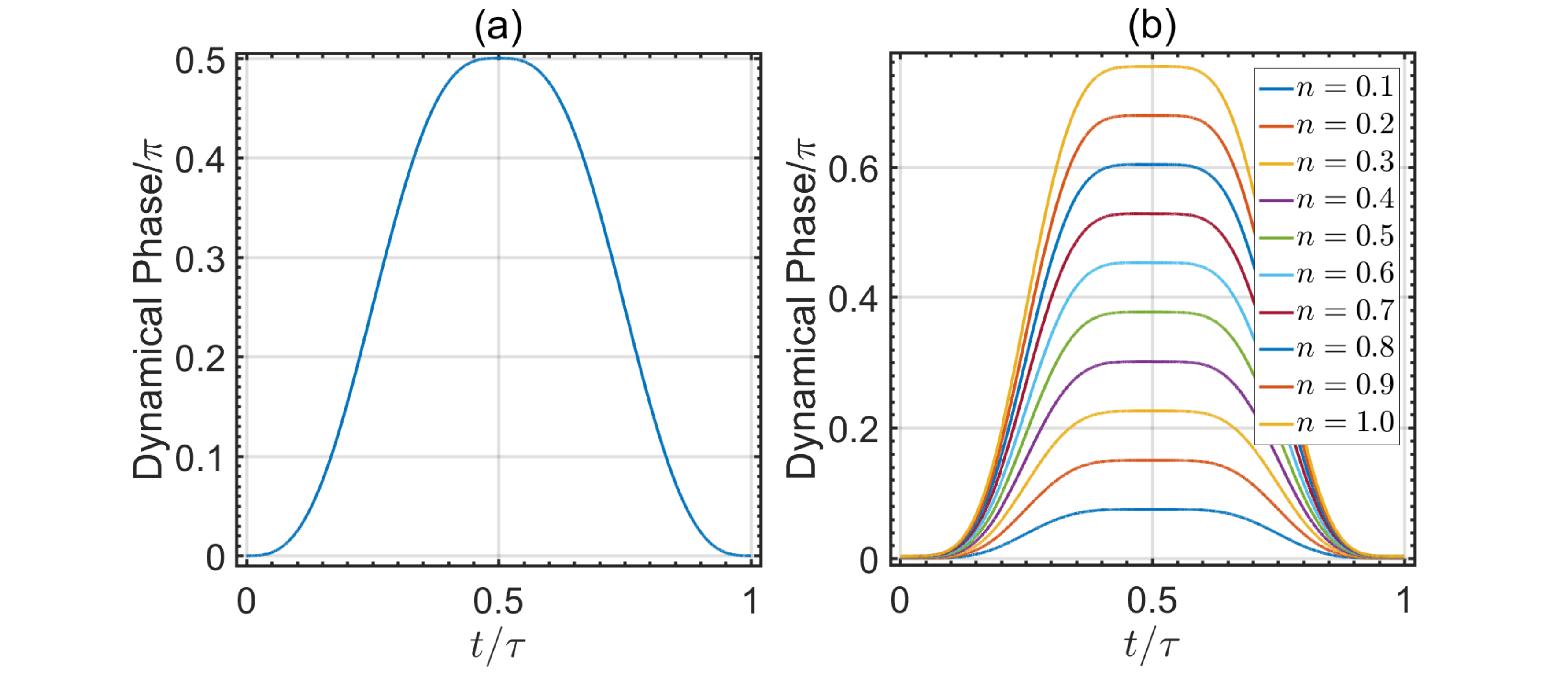}
	\caption{(a)[(b)] Dynamical phase of the scheme in Sec.~\ref{s4.1}~(Sec.~\ref{slogicalqubit})}
	\label{f009}
\end{figure}

\subsection{Numerical results}
In Fig.~\ref{f009}, we plot the dynamical phase of Sec.~\ref{s4.1} and Sec.~\ref{slogicalqubit} numerically, which also shows the dynamical phase equals zero finally.

\section{Derivations of $q_{s}$}\label{s11}
Consider the affect of static systematic error, the initial state of system is in $|\psi_{0}(0)\rangle$, and the unperturbed evolution operator is $\hat{\mathcal{U}}(\upsilon,t)=|\psi_{0}(\upsilon)\rangle\langle\psi_{0}(t)|+|\psi_{\bot}(\upsilon)\rangle\langle\psi_{\bot}(t)|$.
Then $|\psi(\tau/2)\rangle=|\psi_{0}(\tau/2)\rangle-i\varepsilon\int_{0}^{\tau/2}dt\hat{\mathcal{U}}_{0}(\tau/2,t)\hat{\mathcal{H}}_{\rm eff}(t)|\psi_{0}(t)\rangle-\varepsilon^{2}\int_{0}^{\tau/2}dt\int_{0}^{t}dt'\hat{\mathcal{U}}_{0}(\tau/2,t)\hat{\mathcal{H}}_{\rm eff}(t)\hat{\mathcal{U}}_{0}(t,t')\hat{\mathcal{H}}_{\rm eff}(t')|\psi_{0}(t')\rangle$, where we keep it to the second order ignoring the higher order of $\varepsilon$. And $|\psi_{0}(t)\rangle$ and $\hat{\mathcal{U}}_{0}$ denote the unperturbed solution and evolution operator, respectively. Then the fidelity is defined as: $P=\left|\langle\psi_{0}(\tau/2)|\psi(\tau/2)\rangle\right|^{2}=1-\varepsilon^{2}\left|\int_{0}^{\tau/2}dt\langle\psi_{\bot}(t)|\hat{\mathcal{H}}_{\rm eff}|\psi_{0}(t)\rangle\right|^{2}$. The systematic-error sensitivity is defined as $q_{s}=-\frac{1}{2}\frac{\partial^{2}P}{\partial\varepsilon^{2}}|_{\varepsilon=0}=\left|\int_{0}^{\tau/2}dt\langle\psi_{\bot}(t)|\hat{\mathcal{H}}_{\rm eff}|\psi_{0}(t)\rangle\right|^{2}$. Combining the Eqs.~(\ref{e20j}),~(\ref{e23}) and (\ref{e25j}), we further get the expression of $q_{s}$ as
\begin{eqnarray}\label{e34j}
q_{s}&=&\left|\int_{0}^{\tau/2}dt\langle\psi_{\bot}(t)|\hat{\mathcal{H}}_{\rm eff}|\psi_{0}(t)\rangle\right|^{2}\cr\cr
&=&\frac{1}{4}\left|\int_{0}^{\tau/2}dt[-ie^{-i\gamma}\dot{\gamma}\cos\Theta\sin\Theta+e^{-i\gamma}\dot{\Theta}]\right|^{2}\cr\cr
&=&\frac{1}{4}\left|\int_{0}^{\tau/2}dt\left[-e^{-i\gamma}\frac{d}{dt}(\cos\Theta\sin\Theta)+e^{-i\gamma}\dot{\Theta}\right]\right|^{2}\cr\cr
&=&\left|\int_{0}^{\tau/2}dte^{-i\gamma}\dot{\Theta}\sin^{2}\Theta\right|^{2}
\end{eqnarray}
Where we have used the boundary condition of $\Theta(0)=0$, $\Theta(\tau/2)=\pi$ in the derivation of the above formula.

	\bibliography{Refsv1}

\begin{thebibliography}{129}
\bibitem [{not()}]{note}%
  \BibitemOpen
  \href@noop {} {\bibinfo  {journal} {For the stark shifts
  $\frac{\Omega'^{2}}{4\Delta}|\mathcal{B}\rangle\langle
  \mathcal{B}|+\frac{\Omega_{C}^{2}}{4\Delta}|R\rangle\langle R|$, we can
  introduce an auxiliary state $|p'\rangle$, making the laser pulses to
  generate the couplings
  $\frac{1}{2}[e^{-i\Delta'_{t}}\Omega''e^{i\phi_{B}}|\mathcal{B}\rangle\langle
  p'|+e^{-i\Delta''_{t}}\Omega'_{C}|R\rangle\langle p|]+ {\rm H.c.}$, where
  $\Omega'^{2}/\Delta=\Omega''^{2}/\Delta'$,
  $\Omega_{C}^{2}/\Delta=\Omega_{C}'^{2}/\Delta''$, which leads to the opposite
  stark shifts $-[\frac{\Omega'^{2}}{4\Delta}|\mathcal{B}\rangle\langle
  \mathcal{B}|+\frac{\Omega_{C}^{2}}{4\Delta}|R\rangle\langle R|]$ with the
  condition of large detuning approximations, thus the stark shifts can be
  cancelled}\ }\BibitemShut {NoStop}%
\end{thebibliography}

\end{document}